\documentclass[12pt]{iopart}

\usepackage{graphicx}
\usepackage{acronym}

\begin{document}

\newacro{AUG}[AUG]{ASDEX Upgrade}
\newacro{CXRS}[CXRS]{charge exchange recombination spectroscopy}
\newacro{BES}[BES]{beam emission spectroscopy}

\newacro{pe}[$p\rm{_e}$]{electron pressure}
\newacro{DBP}[$D_{\omega}$]{distribution of the observed intensity}
\newacro{ne}[$n\rm{_e}$]{electron density}
\newacro{Te}[$T\rm{_e}$]{electron temperature}
\newacro{fIF}[$f\rm{_{IF}}$]{IF bandwidth}
\newacro{Trad}[$T\rm{_{rad}}$]{radiation temperature}
\newacro{Ti}[$T\rm{_i}$]{ion temperature}
\newacro{DCN}[DCN]{deuterium cyanide}
\newacro{ECE}[ECE]{electron cyclotron emission}
\newacro{ECFM}[ECFM]{electron cyclotron forward modeling}
\newacro{ECEI}[ECE-I]{ECE-imaging}
\newacro{IF}[IF]{intermediate frequency}
\newacro{ICRH}[ICRH]{ion cyclotron resonance heating}
\newacro{ECRH}[ECRH]{electron cyclotron resonance heating}
\newacro{P_ECRH}[$P\rm{_{ECRH}}$]{ECRH power}
\newacro{OH}[OH]{ohmic heating}
\newacro{P_OH}[$P\rm{_{OH}}$]{ohmic heating power}
\newacro{PSL}[PSL]{passive stabilization loop}
\newacro{P_NET}[$P\rm{_{NET}}$]{applied net heating power}

\newacro{LIB}[LIB]{lithium beam}
\newacro{LSQ}[LSQ]{least square}
\newacro{LFS}[LFS]{low field side}
\newacro{HFS}[HFS]{high field side}
\newacro{RFA}[RFA]{resonant field amplification}

\newacro{LCFS}[LCFS]{last closed flux surface}
\newacro{ELM}[ELM]{edge localized mode}
\newacro{H-mode}[H-mode]{high confinement mode}
\newacro{WMHD}[$W\rm{_{MHD}}$]{plasma energy}

\newacro{MP}[MP]{magnetic perturbation}
\newacro{IDA}[IDA]{integrated data analysis}
\newacro{L-mode}[L-mode]{low confinement mode}
\newacro{NBI}[NBI]{neutral beam injection}
\newacro{SOL}[SOL]{scrape off layer}
\newacro{LOS}[LOS]{lines of sight}
\newacro{MHD}[MHD]{magnetohydrodynamics}
\newacro{MAST}[MAST]{}
\newacro{DIII-D}[DIII-D]{}
\newacro{KSTAR}[KSTAR]{}
\newacro{SFL}[SFL]{straight field line}

\newacro{EAST}[EAST]{}
\newacro{JET}[JET]{}

\newacro{TS}[TS]{Thomson scattering}

\title[\small Plasma response measurements of MPs using ECE and comparisons to 3D ideal MHD]{Plasma response measurements of external magnetic perturbations using electron cyclotron emission and comparisons to 3D ideal MHD equilibrium}

\author{M. Willensdorfer$^1$\footnote{matthias.willensdorfer@ipp.mpg.de}, S.S. Denk$^{1,2}$,  E. Strumberger$^1$, W. Suttrop$^1$, B. Vanovac$^3$, D. Brida$^{1,2}$, M. Cavedon$^{1,2}$, I. Classen$^3$, M. Dunne$^1$, S. Fietz$^1$, R. Fischer$^1$, A. Kirk$^4$, F.M. Laggner$^5$, Y. Q. Liu$^4$, T. Odstr\v{c}il$^{1,2}$, D.A. Ryan$^{6,4}$, E. Viezzer$^1$, H. Zohm$^1$, I.C. Luhmann$^7$,  the ASDEX Upgrade Team and the EUROfusion MST1 Team$^*$}

\address{ $^1$ Max Planck Institute for Plasma Physics, 85748 Garching, Germany \\
$^2$ Physik-Department E28, Technische Universit\"at M\"unchen, 85748 Garching, Germany \\
$^3$ FOM-Institute DIFFER, Dutch Institute for Fundamental Energy Research \\
$^4$ CCFE, Culham Science Centre, Abingdon, Oxon, OX14 3DB, UK \\
$^5$ Institute of Applied Physics, TU Wien, Fusion@\"OAW \\
$^6$York Plasma Institute, Department of Physics, University of York, Heslington, York,YO10 5DQ, UK \\
$^7$ University of California at Davis, Davis, CA 95616, USA \\
$^*$ See http://www.euro-fusionscipub.org/mst1    \vspace{-8pt}
 }


\begin{abstract}
The plasma response from an external $n=2$ \acl{MP} field in ASDEX Upgrade  has been measured using mainly \ac{ECE} diagnostics and a rigid rotating field.
To interpret \ac{ECE} and \ac{ECEI} measurements accurately, forward modeling of the radiation transport has been combined with ray tracing.
The measured data is compared to synthetic \ac{ECE} data generated from a 3D ideal \ac{MHD} equilibrium calculated by VMEC. 

The measured amplitudes of the helical displacement around the low field side midplane are in reasonable agreement with the one from the synthetic VMEC diagnostics. Both exceed the prediction from the vacuum field calculations and indicate the presence of a kink response at the edge, which amplifies the perturbation. 
 VMEC and MARS-F have been used to calculate the properties of this kink mode. The poloidal mode structure of the magnetic perturbation of this kink mode at the edge peaks at poloidal mode numbers  larger than the resonant components $|m|>|nq|$, whereas the poloidal mode structure of its displacement is almost resonant $|m|\approx|nq|$. This is expected from ideal MHD in the proximity of rational surfaces. The displacement measured by \ac{ECEI} confirms this resonant response.

\end{abstract}

\maketitle

\acresetall

\section{Introduction}

The usage of non-axisymmetric external \ac{MP}-fields is one method, among others, to suppress \acp{ELM}~\cite{Evans:2004} or to mitigate them~\cite{Liang:2007}. It is utilized in several devices like ASDEX Upgrade~\cite{Suttrop:2011a}, DIII-D~\cite{Evans:2008}, EAST~\cite{Sun:2015}, JET~\cite{Liang:2007}, KSTAR~\cite{Jeon:2012}, \acs{MAST}~\cite{Kirk:2013}.  These experiments have shown that \ac{ELM} mitigation and \ac{ELM} suppression are achievable over a wide range of collisionalities $\nu^\star$. 

At ASDEX Upgrade \ac{ELM} mitigation using external \acp{MP} has been achieved at  high plasma densities ($n_{edge}/n_{GW} > 0.65$ corresponding to $\nu^\star>1.2$)\cite{Suttrop:2011a, Suttrop:2014} and, more recently, also at low pedestal collisionality $\nu^\star$ ($\nu^\star<0.4$)~\cite{Suttrop:2014,Kirk:2015}. Although large type-I \acp{ELM} disappear, small \acp{ELM} with frequencies up to $1\ \rm{kHz}$ remain in both $\nu^\star$ windows. This is different to DIII-D experiments, where \ac{ELM} suppression with quiescent divertor signals has been achieved.  Since the type of the remaining \acp{ELM} during the \ac{MP} phase, especially at low $\nu^\star$, is unclear, we refer to this suppression of large type-I \ac{ELM}  as \ac{ELM} mitigation. 
 
 The \ac{ELM} mitigation at low $\nu^\star$ is accompanied with a decrease of density, the so-called density pump-out \cite{Manolo:2013}.  This is also in-line with  \ac{ELM} mitigation experiments in other devices. It is also observed for \ac{ELM} suppression in DIII-D, which indicates a similar underlying physical mechanism for the increased particle transport. More comprehensive experimental studies in ASDEX Upgrade~\cite{Suttrop:2014, Kirk:2015}, DIII-D~\cite{Paz-Soldan:2015} and MAST~\cite{Kirk:2013} show that the degree of \ac{ELM} mitigation and density pump-out depends on the poloidal spectrum of the external magnetic perturbation.
Moreover, the optimum applied poloidal spectrum for \ac{ELM} mitigation does not show a maximum of the pitch-aligned magnetic field  component. Instead, it is aligned with the mode at the edge that is most strongly amplified by the plasma~\cite{Kirk:2013, Paz-Soldan:2015}, as calculated with \ac{MHD} response models like MARS-F~\cite{Liu:2000}, JOREK~\cite{Orain:2015} and VMEC~\cite{Strumberger:2014}.
These \ac{MHD} calculations also suggest that this plasma response is a composition of pressure-driven kink modes and a current driven response referred as the low-$n$ peeling response, which can couple to resonant components~\cite{Ryan:2015, Orain:2016}. Their individual contributions vary with the applied poloidal mode spectrum. 
The kink mode is located around the \ac{LFS} midplane~\cite{Lanctot:2011, Moyer:2012}, whereas the  peeling response is predicted to peak around the X-Point and the top of the plasma~\cite{Shafer:2014,Liu:2016}. 
The poloidal mode structures of both are dominated by poloidal mode numbers $m$ larger than the resonant components $|m|>|nq|$. 
Further experimental investigations indicate that this X-point peeling response, rather than the kink mode at the \ac{LFS},  causes  the \ac{ELM} mitigation and the density pump-out~\cite{Kirk:2013,Kirk:2015,Paz-Soldan:2015}. 

In principle, the kink response can amplify the external magnetic perturbation, which results in a pronounced non-axisymmetric displacement of inter alia the \ac{LCFS} at the \ac{LFS}~\cite{Chapman:2012}. Although the kink response seems to play a minor role in \ac{ELM} mitigation, the effect of this distortion on \ac{ELM} stability is not completely clear. Moreover, this displacement can also lead to unwanted effects of the position control system on the plasma like unintended movements of the plasma~\cite{Chapman:2014b}. Hence, it is important to characterize it and to compare it to \ac{MHD} codes.

 
 In this paper, we describe a method to measure the non-axisymmetric flux surface displacement using toroidally localized \ac{ECE} measurements and rigid rotating $n=2$  \ac{MP}-field. A similar method has already been used in Refs.~\cite{Tobias:2013a, Tobias:2013b}. The kink response of DIII-D plasmas has been compared to IPEC calculations~\cite{Park:2007}. We extended this method using forward modeling of the electron cyclotron radiation transport from Ref.~\cite{Rathgeber:2013} and additional ray tracing, which provide the accuracy needed to study the kink response at the edge. To allow comparisons with 3D ideal \ac{MHD} equilibrium calculations from VMEC~\cite{Hirshman:1986}, we developed synthetic VMEC diagnostics. In case of synthetic \ac{ECE} diagnostics, we combined the forward modeling and the 3D equilibrium from VMEC. The amplitude of the plasma surface displacement and its poloidal mode structure are quantitatively compared. Additional profile diagnostics like the \ac{LIB} and \ac{CXRS} as well as MARS-F calculations complement the comparison.

This paper is organized as follows. Section~\ref{sec:setup} describes the measurement principle, the experimental setup, the magnetic perturbation coil setup and the diagnostic tools with  focus on \ac{ECE} diagnostics. The forward modeling of the \ac{ECE} systems is  presented in Section~\ref{sec:ECEmodel}. VMEC calculations and the implementation of synthetic diagnostics are explained in Section~\ref{sec:VMEC}. Section~\ref{sec:amplitude} and \ref{sec:poloidal} present the comparison of the amplitude and of the poloidal mode spectrum, respectively. 
Conclusions and a summary are given in Section~\ref{sec:conclusions}.

\section{Experimental setup}
\label{sec:setup} 
 \begin{figure*}[htc]
   \centering
 \includegraphics[width=1.0\textwidth]{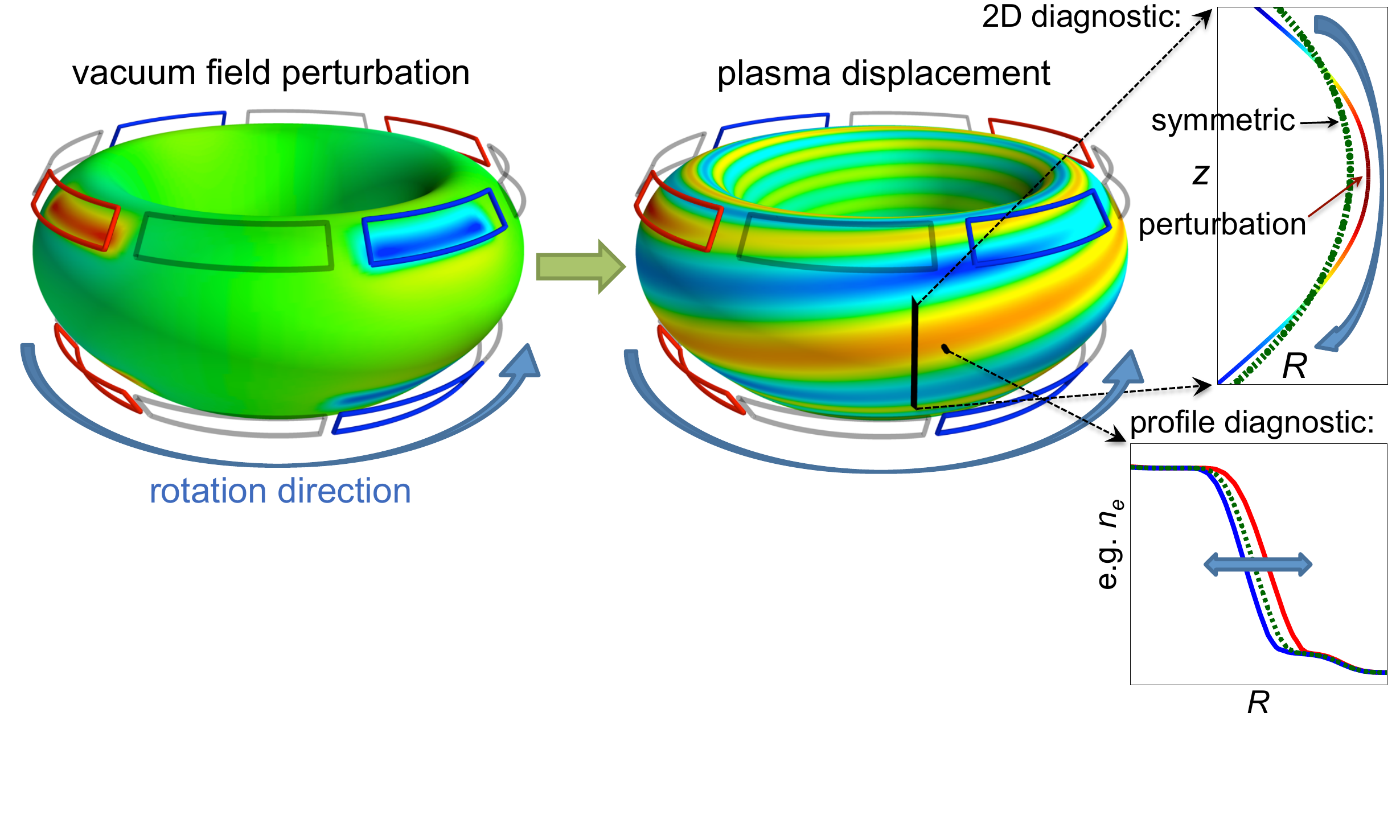} 
 \caption{Cartoon of measuring the plasma displacement using a rigid rotating external \ac{MP}-field. (left) external 'saddle' coils produce a \ac{MP} of the vacuum field. (middle) this \ac{MP}-field causes a perturbation of the flux surfaces (only \acs{LCFS} is shown). The color scaling of the surface plots indicates the magnetic perturbation $B_r$ (left) and the surface perturbation $\xi_r$ (middle) into direction normal to the \acs{LCFS}. Red indicates a perturbation pointing outwards, blue inwards and green none.  A rotation  of the external \ac{MP}-field (left) results in a rotation of the displacement (middle),  which can be  measured by an imaging system (right, top) or profile diagnostics (right, bottom). The rotation direction of the rigid rotation (in positive toroidal angle $\phi$) is  indicated by blue arrows.}
\label{fig:rigidprinciple}
\end{figure*}

The measurement principle is based on external saddle coils, which produce non-axisymmetric \acp{MP} of the vacuum field. The result is a nearly pitch-aligned non-axisymmetric distortion of the flux surfaces, which is static to the \ac{MP}-field. The main idea is that a rotation of this external \ac{MP}-field leads to a rotation of the displacement (illustrated in Figure \ref{fig:rigidprinciple}). This rotating distortion is then measured by profile and/or imaging diagnostics~\cite{Moyer:2012}. The rotating distortion should appear in profile diagnostics as radially varying displacement. Their high spatial resolution can be used to accurately measure the amplitude of the distortion \cite{Fischer:2012, Fuchs:2014}.
From the imaging diagnostics, we can gain  information about the alignment of the distortion by inspecting the poloidal phase of the plasma response as a function of the continually varied toroidal phase of the external \ac{MP}-field.
The present experiment was made with the toroidal magnetic field and plasma current pointing in opposite directions, hence negative safety factor q.
Consequently, an imaging system should detect a poloidally downward propagating structure, if the rotation is in positive toroidal direction (counterclockwise in the cartoon). The rotation directions are indicated by blue arrows in Fig.~\ref{fig:rigidprinciple}. 

In this paper, we  focus on \ac{ECE} diagnostics for profile and imaging measurements. They are ideal to track changes of the flux surfaces, since they are able to deliver the \ac{Te} with high temporal resolution. Due to the very large electron heat diffusivity along the magnetic field lines, \ac{Te} is essentially constant on flux surfaces.

\subsection{\ac{MP}-coils and edge diagnostics}

ASDEX Upgrade is equipped with 16 \ac{MP}-coils, which are arranged in two poloidally separated rows and each has eight toroidally equidistant coils~\cite{Suttrop:2009a} (shown in Fig.~\ref{fig:rigidprinciple}). 
This allows us to apply an \ac{MP}-field with a toroidal mode number $n$ of 1, 2 and 4. A newly installed power supply system enables us to rotate the \ac{MP}-field of the two coil sets separately~\cite{Teschke:2014}. Thus, it is possible to employ either a differential rotation (sets in opposite direction) or a rigid rotation (both sets in same direction)  using $n=1,2$ with frequencies up to several $100\ \rm{Hz}$~\cite{Teschke:2014}. Because of the \ac{PSL} conductors in ASDEX Upgrade, fast rotating \ac{MP}-fields are attenuated and delayed by image currents~\cite{Suttrop:2009b}. To avoid significant attenuation, we applied a low frequency of $0.5\ \rm{Hz}$ for the $n=2$ rigid rotation~\cite{Suttrop:2016}. The estimated attenuation is not more than $10\%$. We used even parity configuration for the rigid rotation, which means that the differential phase angle between the \ac{MP}-field of the upper and lower coil set $\Delta \phi \rm{_{ul}}$ is $0^{\circ}$ \cite{Suttrop:2011b}. Since there are 8 saddle coils in each row, the toroidal mode spectrum of the external \ac{MP}-field exhibits a dominant $n=2$ and weak $n=6,10,14,\dots$ components but no other additional side bands. The intrinsic error field for $n=2$ is assumed to be small, because: First, no global density perturbation have been observed during the $n=2$ rigid rotation and second, dedicated measurements of the $n=1$ error field also indicates only a small $n=2$ component (no ellipse in Fig.~4 in Ref.~\cite{Maraschek:2013}).

 \begin{figure*}[htc]
   \centering
 \includegraphics[width=1.0\textwidth]{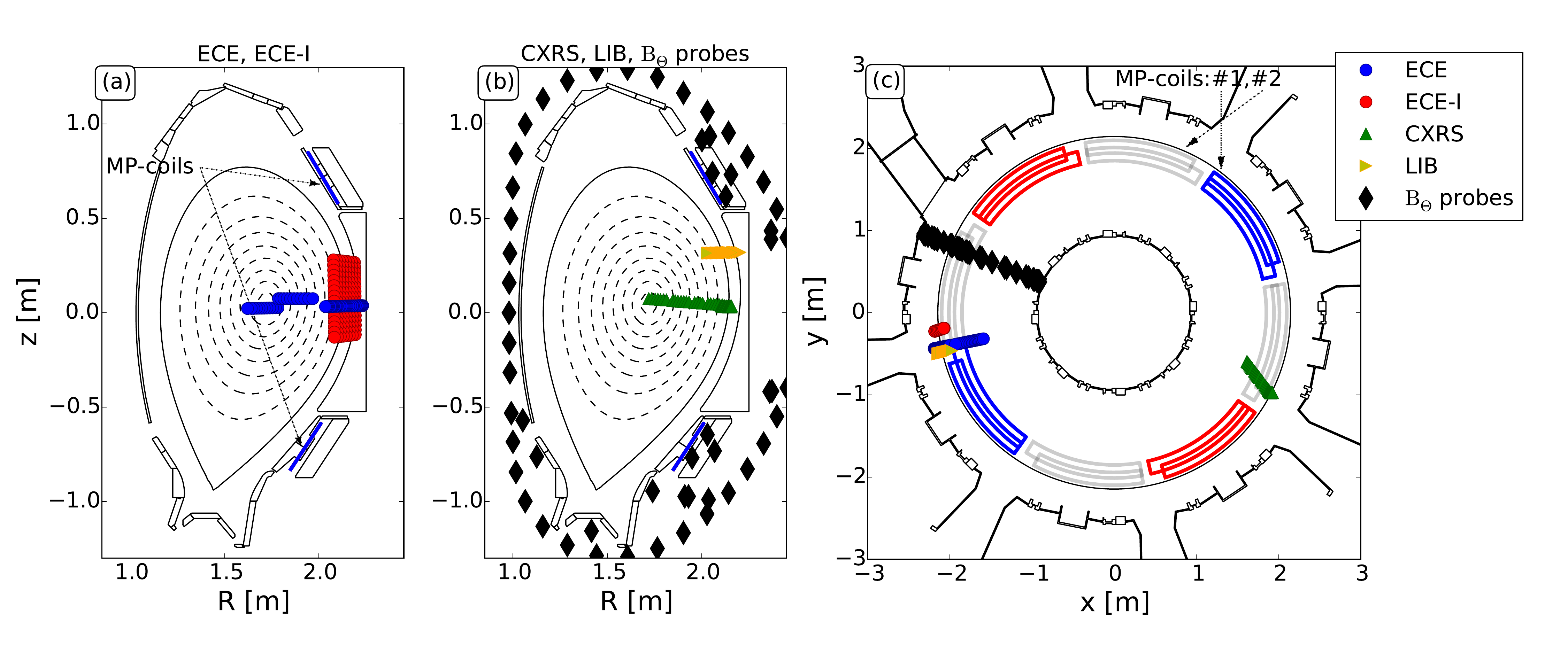} 
 \caption{ Overview of the used diagnostics and the \ac{MP}-coils. (a) and (b) are poloidal cross-section showing \ac{ECE}, \acs{ECEI} and \ac{CXRS}, \ac{LIB}, $B_{\Theta}$ probes, respectively. (c) top view. Dotted arrows indicate the poloidal (a) and the toroidal position (c) of the \ac{MP}-coils. The coil numeration  of each coil-set is shown in (c). The colors of the \ac{MP}-coils illustrates an $n=2$ perturbation.}
\label{fig:diags}
\end{figure*} 

To obtain the edge displacement accurately, several high resolution edge diagnostics are in use. Figure~\ref{fig:diags} shows the poloidal (Fig.~\ref{fig:diags}(a),(b))  and toroidal (Fig.~\ref{fig:diags}(b)) positions of  \ac{CXRS} \cite{Viezzer:2012}, \ac{LIB}~\cite{Willensdorfer:2012, Willensdorfer:2014}, \ac{ECE} and \ac{ECEI}, which measure \ac{Ti},  \ac{ne} and \ac{Te}, respectively. Additionally, the magnetic probes used for the equilibrium reconstruction and plasma position control are shown. 

As mentioned previously, we mainly use \ac{ECE} measurements to track 3D distortions of the flux surfaces.
To obtain reliable edge profiles of \ac{Te} from \ac{ECE}, it is necessary to forward model the electron cyclotron radiation transport~\cite{Rathgeber:2013}. 
The 1D-profile \ac{ECE}  diagnostic (blue circles in Fig.~\ref{fig:diags}) uses a heterodyne radiometer with 60 channels and a sampling rate of $1\ \rm{MHz}$ to measure the second harmonic extraordinary mode (X2). At the standard magnetic toroidal field configuration of $B\rm{_T}\approx-2.5\ \rm{T}$, 36 channels cover the edge region with a spatial resolution of about $5\ \rm{mm}$. This spatial resolution is set by the frequency spacing between the channels, the  \ac{IF} bandwidth of $300\ \rm{MHz}$ of each channel and the additional broadening due to electron cyclotron radiation transport effects like Doppler broadening. The remaining 24 channels are used to measure the core using a frequency spacing of $\approx1\ \rm{GHz}$ and an \ac{fIF} $600\ \rm{MHz}$ resulting in a spatial resolution of about $12\ \rm{mm}$. The 1D-profile \ac{ECE} system is calibrated absolutely \cite{Carli:1974,Hartfuss:1997}, whereas the \ac{ECEI} system relies on a cross calibration. 

The used \ac{ECEI} system has 128 channels (red circles in Fig.~\ref{fig:diags}) with a temporal resolution of $200\ \rm{kHz}$~\cite{Classen:2010}. It was configured to cover the plasma edge using X2. It has 16 rows with a vertical spacing of $\approx 25\ \rm{mm}$ and 8 channels in each row. The frequency spacing is $800\ \rm{MHz}$, whereas \ac{fIF} is $700\ \rm{MHz}$. The resulting radial spatial resolution is around $15\ \rm{mm}$ at the edge. The advantage of \ac{ECEI}  is that the vertical distribution of the channels allows us to resolve poloidal structures. Because of a recent extension to a quasi 3D system \cite{Classen:2014}, the toroidal angle between the geometrical \ac{LOS} of the \ac{ECEI}  system and the toroidal field is oblique. This complicates the interpretation of the \ac{ECEI} system and it is necessary to forward model the \ac{ECEI}. This is treated in detail in section~\ref{sec:ECEmodel}.

Measuring the 3D displacement using toroidally localized diagnostics and a rigid rotating \ac{MP}-field is based on  two assumptions: First, the measured plasma parameters are constant on the perturbed flux surfaces and second, global plasma parameters, like core temperature and density, do not change significantly during the rigid rotation. 
The validity of both assumptions is justified in the following section.

\subsection{Discharge}

\begin{figure}[htc]
   \centering
 \includegraphics[width=0.5\textwidth]{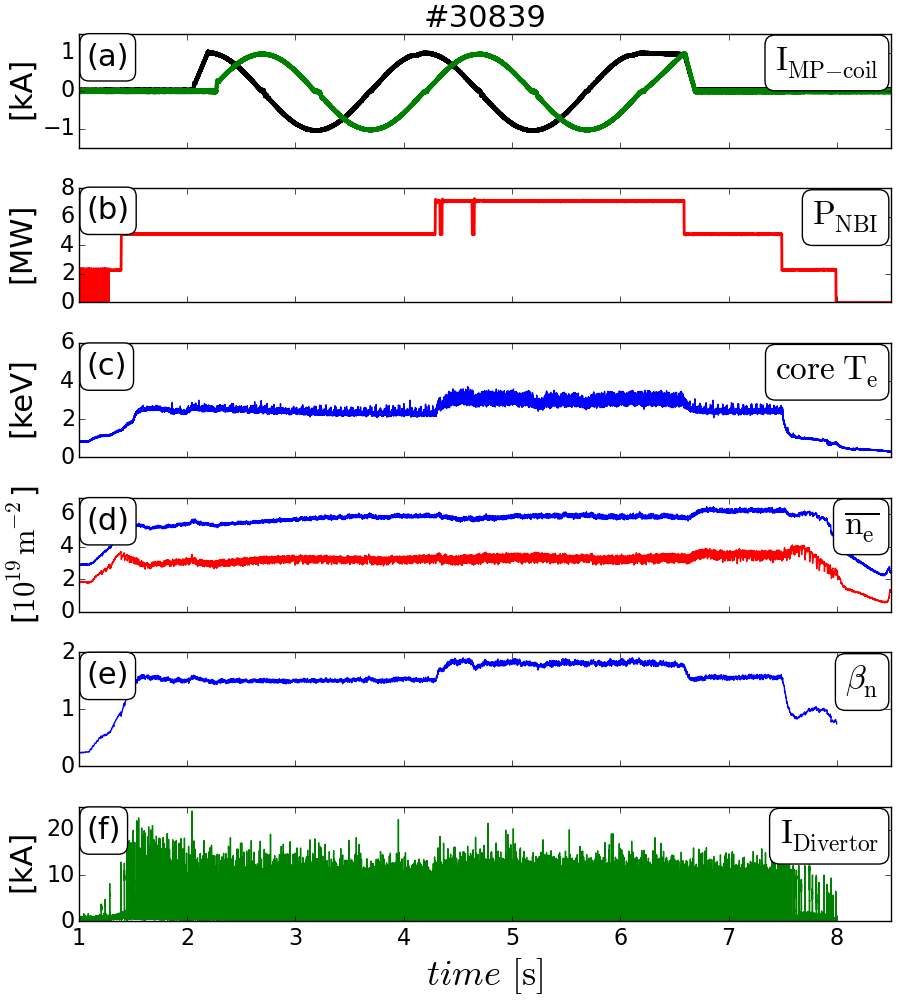} 
 \caption{Discharge overview: (a) coil current of the first (black) and second (green) \ac{MP}-coil from the upper coil-set (see Fig.~\ref{fig:diags}(c)) to indicate the time period of the rigid rotation, (b) heating power, (c) core electron temperature, (d) line integrated density from a core (blue) and an edge chord (red), (e) normalized beta $\beta_N$ and (f) divertor current. The rigid rotation does not alter global plasma parameters.}
\label{fig:discharge}
\end{figure}

The presented experiment at ASDEX Upgrade was done at a plasma current of $I_P = 800\ \rm{kA}$ and a toroidal field of $B\rm{_T} = -2.5\ \rm{T}$ (direction of $B\rm{_T} $ is clockwise). In this configuration, the direction of the ion $\nabla B$ drift is towards the X-point and the edge safety factor amounts to $q\rm{_{95}} \approx -5.4 $. Figure~\ref{fig:discharge} shows the time traces of global plasma parameters during the application of the \ac{MP}-coils. The rotation was performed with a frequency of $0.5\;\rm{Hz}$ indicated by the \ac{MP}-coil currents in Fig.~\ref{fig:discharge}(a). Two periods in positive toroidal direction were performed and in-between the \ac{NBI} power was stepped from  $5$ to $7.5~\rm{MW}$. Within one \ac{NBI} step, the density and temperature do not vary more than $6\%$ in the core (Fig.~\ref{fig:discharge}(c,d)). The normalized beta $\beta_N$ amounts to $\approx1.7$ and $\approx2$  in the second \ac{NBI} power step (see Fig.~\ref{fig:discharge}(e)). The application of this \ac{MP}-coil configuration does not significantly affect the \ac{ELM} behavior as seen in the divertor current (Fig.~\ref{fig:discharge}(f)). \ac{ELM} mitigation is also not expected for these plasma parameters with an even parity configuration ($\Delta \phi \rm{_{ul}}\approx0^{\circ}$). The optimum phase angle for \ac{ELM} mitigation for this 'high $B\rm{_T}$' and 'high $q$' case scenario is, according to MARS-F~\cite{Liu:2000} calculations,  $\Delta \phi \rm{_{ul}}\approx-90^{\circ}$ (Fig.~11 in Ref.~\cite{Liu:2016}).
 
It is clearly seen in Figure~\ref{fig:discharge} that core densities and temperatures do not change significantly during the rigid rotation (less than $6\%$). This confirms the assumption of constant global plasma parameters. 
The time traces of Figure~\ref{fig:discharge}  also suggests that the first assumption of constant measured plasma parameters on perturbed flux surfaces in the pedestal region is fulfilled. The breaking up of flux surfaces due to stochastization in the bulk of the pedestal would result in a significant decrease of the temperature and density gradients in the pedestal and, consequently, also of the core temperature and density. Since there is no pronounced drop of these parameters during the switching on of the \ac{MP}-field, a stochastization of the entire gradient region can be ruled out.  Strike point splitting~\cite{Gao:2015} is observed, which indicates a break in axisymmetry. But there is no hint for stochastization within the edge region.
Moreover, we can also assume that ideal \ac{MHD} is applicable in this case, which is discussed in section~\ref{sec:idealMHD}.


\subsection{Edge measurements during rigid rotation}
\label{edgeMeasurements}

\begin{figure}[htc]
   \centering
 \includegraphics[width=0.5\textwidth]{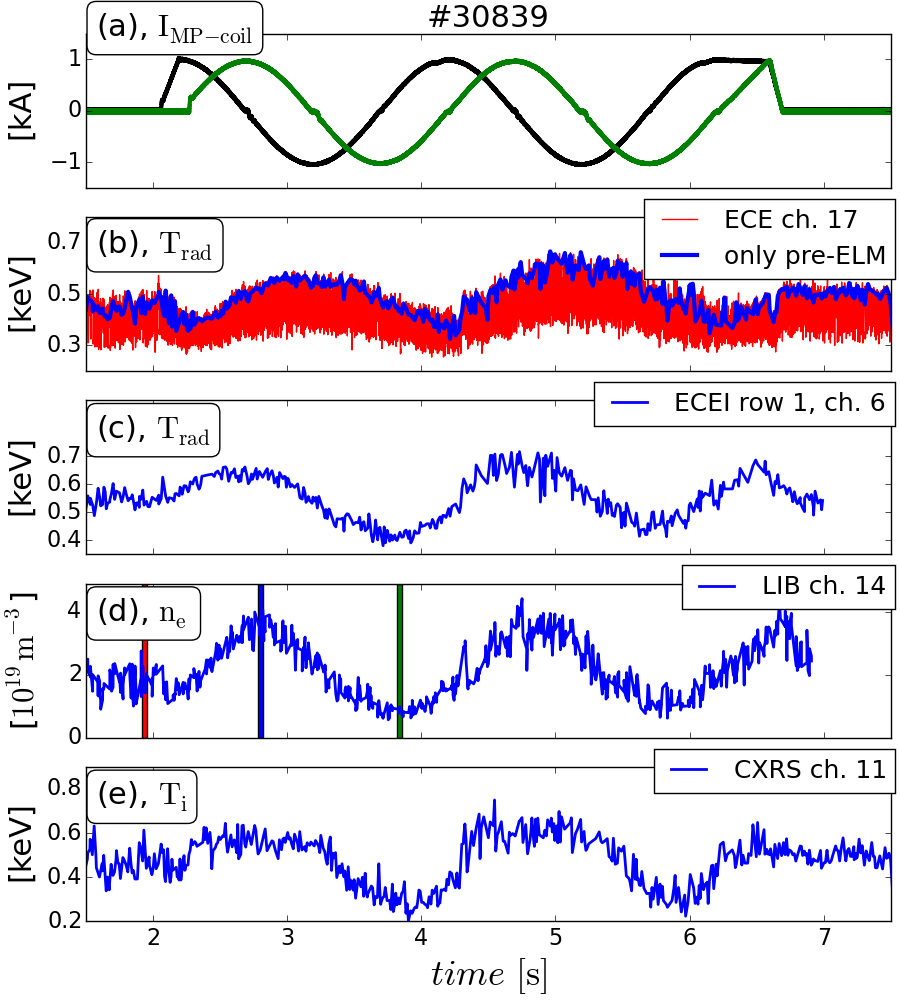} 
 \caption{Edge measurements during the rigid rotation: (a) same coil currents as in Fig.~\ref{fig:discharge}(a) to indicate the rigid rotation, (b) time traces of one edge \ac{ECE} channel using all time points (red) and using pre-\ac{ELM} time points only (blue). Panel (c), (d) and (e) show time traces from one \ac{ECEI} channel, \ac{LIB} and edge \ac{CXRS}, respectively,  using pre-\ac{ELM} time points only. Vertical bars in (d) indicate the time windows used in Fig.~\ref{fig:LIBprofiles}. The edge perturbation is seen in each edge diagnostic.}
\label{fig:edgetraces}
\end{figure}

Although core \ac{ne}, \ac{Te} and \ac{Ti} values are almost constant in time,  every edge profile diagnostic observes a radial position shift due to the rigid rotation of the \ac{MP}-field. 
Time traces from \ac{ECE}, \ac{ECEI}, \ac{LIB} and edge \ac{CXRS} in Fig.~\ref{fig:edgetraces}  show a clear modulation due to the radial shift. To visualize the modulation, we only use  data from pre-\ac{ELM} time points ($50-80\%$ of the \ac{ELM} cycle). Figure~\ref{fig:edgetraces}(b) illustrates \ac{ECE} measurements  using  all time points (red) and  using  pre-\ac{ELM} time points only (blue). 

\begin{figure}[htc]
   \centering
 \includegraphics[width=0.4\textwidth]{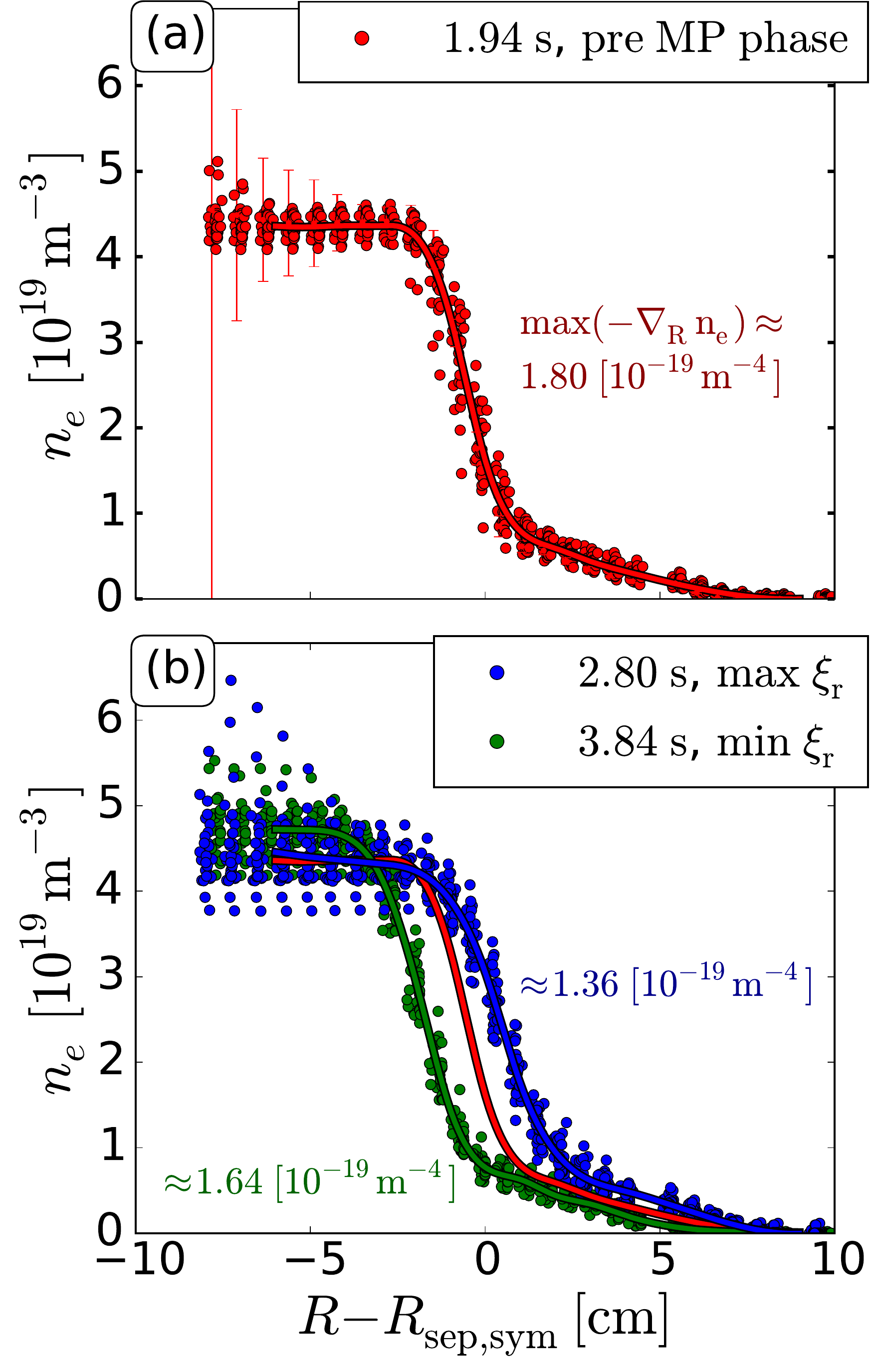} 
 \caption{$\#30839$, edge \ac{ne} profiles from the \ac{LIB} diagnostic versus $R$ relative to the axisymmetric separatrix using only pre-\ac{ELM} time points: (a) profiles before the \ac{MP}-onset (red). The measurement uncertainties from one time point are indicated by error-bars in (a). The maximum density gradient in real space (text inset) are determined from the spline (solid line). (b) same for profiles at the maximum (blue) and  minimum (green) displacement $\xi_r$. For comparison, the fit (red line) from (a) is shown. The analyzed time windows are indicated by vertical bars in Fig.~\ref{fig:edgetraces}.  The gradients are changing during the rotation suggesting a flux surface expansion and compression.}
\label{fig:LIBprofiles}
\end{figure}

To demonstrate that the \ac{MP}-field perturbs the entire edge profile, Fig.~\ref{fig:LIBprofiles} shows  edge  \ac{ne} profiles from \ac{LIB} before the \ac{MP} onset (Fig.~\ref{fig:LIBprofiles}(a) red),  at the maximum displacement (Fig.~\ref{fig:LIBprofiles}(b) blue)  as well as at the minimum displacement (Fig.~\ref{fig:LIBprofiles}(b) green). To account for additional small plasma movement within the analyzed time windows of \rm{$40\ \rm{ms}$}, the profiles are plotted against $R$ relative to the separatrix position determined from the routinely  used axisymmetric equilibrium reconstruction (temporal resolution is $1\ \rm{ms}$)~\cite{McCarthy:2012}. The steep gradient region is well determined by the  \ac{LIB} diagnostics, whereas the pedestal top measurements exhibit relatively large uncertainties (see Fig.~\ref{fig:LIBprofiles}(a)) and, hence, large scatter. This is because of a decreasing sensitivity of the \ac{LIB} diagnostics towards the plasma core~\cite{Willensdorfer:2014}. However, the edge \ac{ne} profiles between the maximum and the minimum displacement show a clear change in the real space gradients, whereas pedestal top values remain, within their uncertainties, the same. This indicates that the \ac{MP}-field induces flux surface expansions and compressions, which depends on the toroidal phase. 

\begin{figure}[htc]
   \centering
 \includegraphics[width=0.5\textwidth]{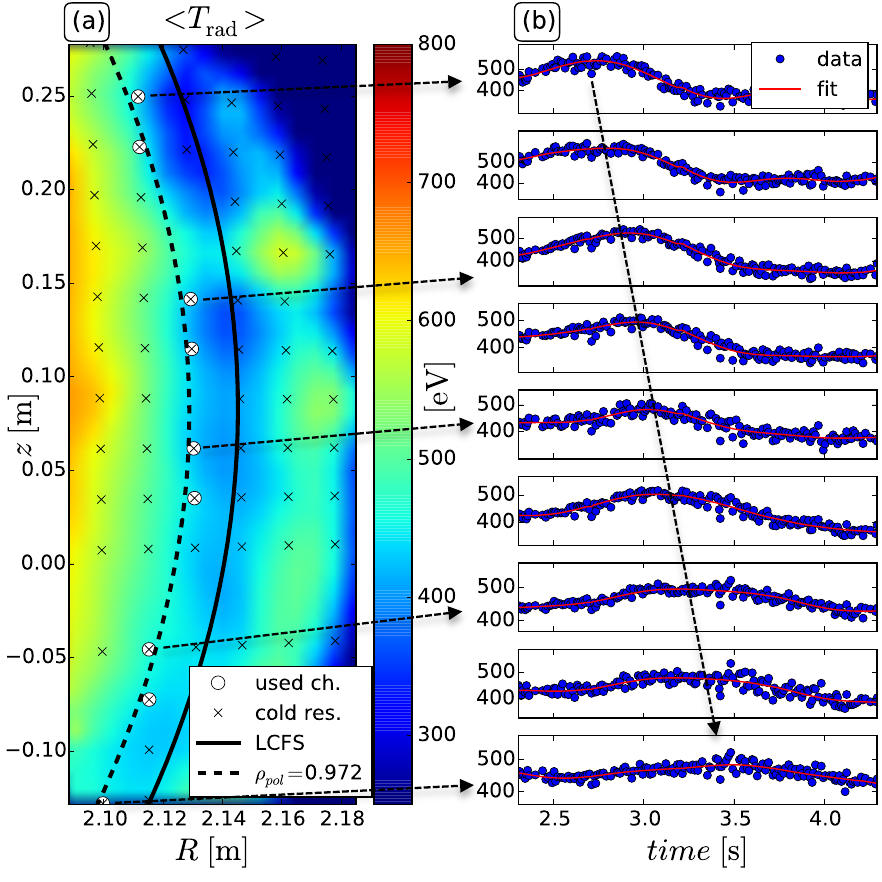} 
 \caption{ (a) The color code shows the mean of radiation temperature $<T\rm{_{rad}}>$ throughout the rigid rotation ($t\approx2.2-4.4\;\rm{s}$) measured by the \ac{ECEI}.  Crosses mark the channel positions. Circles indicate channels close to a flux surface in the pedestal $\rho_{pol}\approx0.972$. The  lines indicate a flux surface in the pedestal $\rho_{pol}\approx0.972$ (dashed) and the last closed flux surface (solid). (b) Time traces of  $T\rm{_{rad}}$ and their corresponding \ac{LSQ} fit from channels marked as circles in (a). The modulation is propagating downwards.}
\label{fig:ECEItraces}
\end{figure}

In addition to the toroidal symmetry breaking, the \ac{MP}-field also adds poloidal structures. This is expected from various  plasma response calculations \cite{Turnbull:2012} and is observed by the \ac{ECEI} system.
Figure \ref{fig:ECEItraces}(a) shows the mean radiation temperature during the rigid rotation using the cold resonance positions for the mapping (details about cold resonance position in Section \ref{sec:ECEmodel}). The solid line indicates the \ac{LCFS} and the dashed line a flux surface within the pedestal region at a normalized poloidal flux of $\rho\rm{_{pol}}\approx0.972$ (the used definition of $\rho\rm{_{pol}}$ is in Ref.~\cite{Fischer:2010}). Time traces using only pre-\ac{ELM} time points of channels along this flux surface and a \ac{LSQ} fit of a sine function including their higher harmonics are shown in Figure \ref{fig:ECEItraces}(b). The modulation is observed in each of these channels.  Furthermore, this modulation is propagating downwards as expected from a \ac{MP}-field rotation in positive toroidal direction (see Fig.~\ref{fig:rigidprinciple} for illustration). 

\ac{ECEI} measurements during the rigid rotation contain valuable information about the flux surface displacement. The perturbation is usually characterized using the Fourier decomposition of its  normal component $\xi \rm{_r}=\xi_a \ e^{i (m \Theta^{\star} - n \phi )}$, where $\xi_a$ is the amplitude, $n$ the toroidal mode number, $\phi$ the toroidal angle, $m$ the poloidal mode number and $\Theta^{\star}$ the \ac{SFL} angle \cite{Dhaeseleer:1991} (see Section \ref{sec:straight}).  $\xi \rm{_r}$  directly measures the displacement between the axisymmetric and 3D equilibrium~\cite{Strumberger:2014}.
 Because of its poloidally and radially localized channels, the \ac{ECEI} diagnostic is able to resolve the poloidal mode numbers $m$. One can obtain $m$ using $m=\frac{\Delta \phi}{\Delta  \Theta^{\star}}$, where $\Delta \phi$ and $\Delta \Theta^{\star}$  are the toroidal phase increment and the corresponding \ac{SFL} angle difference between the various \ac{ECEI} channels.
Therefore, the determination of $m$ depends strongly  on the accuracy of the calculated \ac{SFL} angle. Because of the high shear in the pedestal region, the calculation of $\Theta^{\star}$ is very sensitive to the used flux surface. 
The correct positions and the corresponding flux surfaces of the \ac{ECEI} channels are therefore essential to determine the  poloidal mode structure accurately. 
To provide accurate measurement positions of the \ac{ECEI} channels, we applied forward modeling of the radiation transport, which is described in the next section.

\newpage

\section{Interpretation of ECE}
\label{sec:ECEmodel}

The \ac{ECEI} diagnostic was extended to allow quasi-3D measurements using a second \ac{ECEI} system, which was not in use at the time of this experiment~\cite{Classen:2014}. To enable these quasi-3D measurements, it was necessary to change the \ac{LOS} geometry (see Ref.~\cite{Classen:2014}). This increased the toroidal inclination angle $\angle$ between the \ac{LOS} and the normal to the flux surface or rather the magnetic field line. Therefore, the \ac{LOS} are oblique and not perpendicular anymore, which enhances the Doppler-shift of the observed \ac{ECE} intensities. As a consequence, the position where the measured frequency fulfills the electron cyclotron resonance condition
 (cold resonance position)~\cite{Bornatici:1983}  is not a good approximation for the measurement position or rather the position of the observed \ac{ECE} intensity. The reason for this is that the cold resonance position does not account for kinetic effects like the relativistic mass increase and the Doppler-shift~\cite{Hartfuss:1997}. This section describes the determination of the distribution of the observed \ac{ECE} intensity and its maximum is labelled as 'warm' resonance position ('warm' because kinetic effects are also included, see also Ref.~\cite{Garstka:2001,Marushchenko:2007}). The analysis in this section is done for a time-point prior to the \ac{MP} onset, hence, axisymmetry is assumed.  

\subsection{Definition of the 'warm' resonance position}

\ac{Te} is routinely determined from the \ac{Trad} of the 1D ECE measurements using a forward model within the framework of the \ac{IDA} \cite{Rathgeber:2013, Denk:2016a}. The \ac{Te} profile is varied until the most likely match between the measured and estimated \ac{ECE} intensity within the Bayesian analysis  is found. The \ac{ECE} intensity is calculated by solving the radiation transport equation along the \ac{LOS} \cite{Rathgeber:2013} of the \ac{ECE} diagnostic for given \ac{Te} and \ac{ne} profiles. Because we are mainly interested in the origin of the observed intensity, we only use the module of \ac{IDA}, which solves the radiation transport equation (details in Ref.~\cite{Rathgeber:2013}). For this purpose, $T_e$ and $n_e$ profiles from routine \ac{IDA} evaluation serve as input~\cite{Fischer:2010}.

\begin{figure}[htc]
   \centering
 \includegraphics[width=0.5\textwidth]{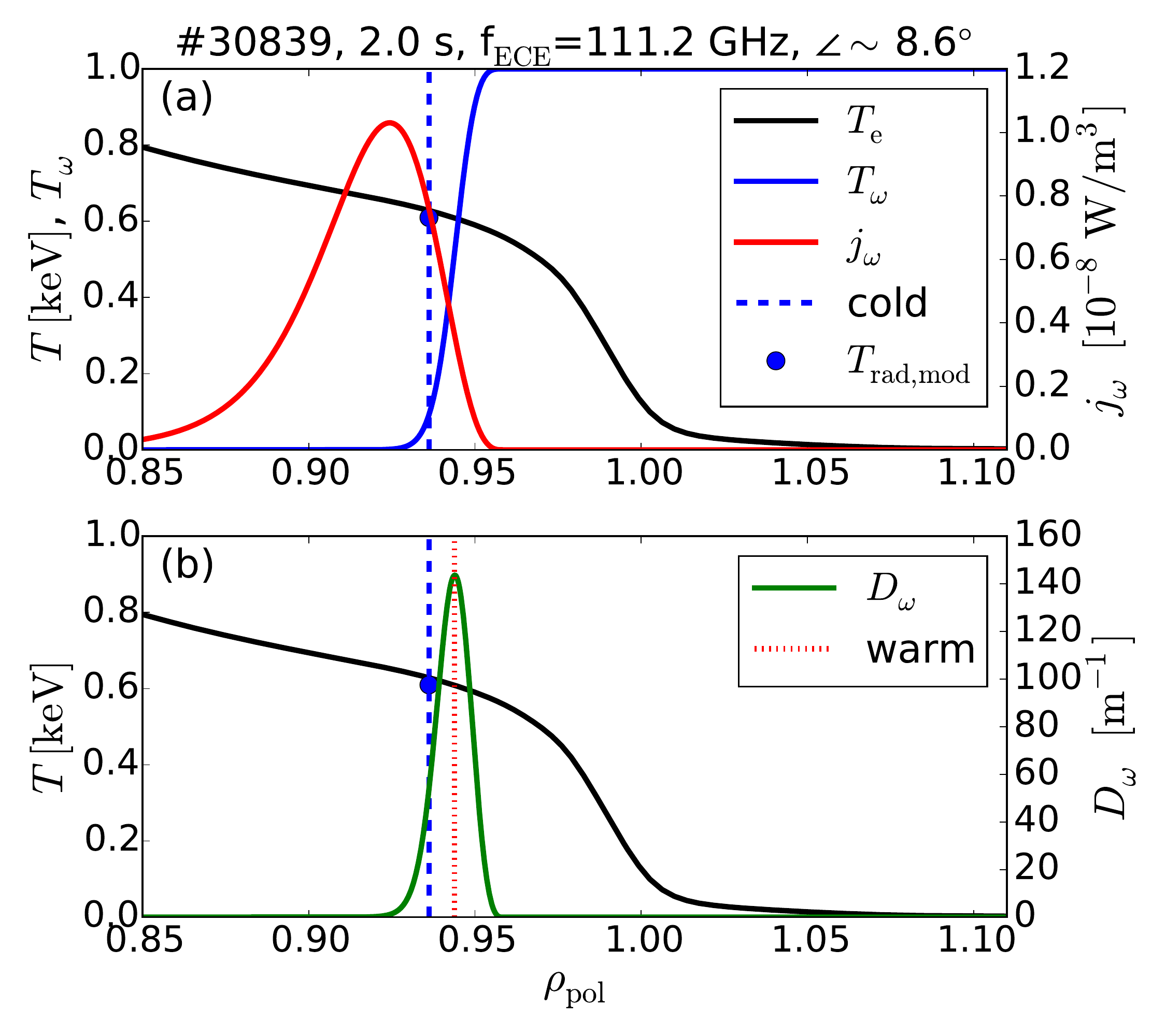} 
 \caption{$\#30839$ at $2.0\ \rm{s}$, (a) emissivity $j_{\omega} $ and the transmittance $T_{\omega}$ are shown  along the ray path mapped on $\rho \rm{_{pol}}$. (b)  The resulting \acl{DBP} (\acs{DBP}) and its maximum, the 'warm' resonance position is indicated by a vertical dotted line. In both frames, the used $T_e$ profile (solid, black) and the cold resonance position (vertical dashed) are shown.}
\label{fig:ECEwarm}
\end{figure}

To account for additional refraction, we extend the modeling by ray tracing (details in Ref.~\cite{Denk:2016a}), which is found to be in good agreement with the TORBEAM code~\cite{Poli:2001}. The ray tracing code calculates the ray path of each channel until the ray hits the wall. Then, the radiation transport equation is solved along the ray path starting from the end of the ray towards the diagnostic antenna. The combination of forward modeling and ray tracing allows us to determine exactly the origin of the observed intensity. The \ac{DBP}~\cite{Denk:2016a} is the normalized product of the emissivity $j_w (s)$ and the transmittance $T_{\omega}(s)$  along the ray path coordinate $s$:
 \begin{equation}
D_{\omega}(s)=\frac{j_w (s)\ T_{\omega}(s)}{ \int j_w (s)\ T_{\omega}(s) ds}
\end{equation}
Figure \ref{fig:ECEwarm}(a) and (b) illustrate $j_w$, $T_{\omega}$ \cite{Denk:2016a} and the resulting \ac{DBP}, respectively, versus $\rho_{pol}$ calculated for one 1D-\ac{ECE} channel  in the pedestal region. The used \ac{Te} profile and the modeled \ac{Trad} ($T_{\rm{rad,mod}}$) value of this channel at its cold resonance position are also shown.
Although we neglect the \ac{IF} bandwidth, the calculated \ac{DBP} is relatively broad, which is attributed to the Doppler and the relativistic broadening. 

To have a single quantity as an approximation for the measurement position of a single \ac{ECE} channel, we use  the maximum of \ac{DBP} labelled as 'warm' resonance position \cite{Sato:1995} (shown in Fig.\ \ref{fig:ECEwarm}). As indicated by the vertical lines in Fig.~\ref{fig:ECEwarm}(b), the 'warm' resonance position can differ from the cold resonance position. This discrepancy originates  from the Doppler effect due to an oblique \ac{LOS} of $\angle\approx8.6^{\circ}$. Although the toroidal inclination angle of the profile \ac{ECE} system amounts  to only $\approx4^{\circ}$ and almost no poloidal inclination angle, additional refraction by the plasma leads to an even larger angle at the cold resonance position. As a result, the Doppler-shift  becomes more important, especially, in the case of the \ac{ECEI} system.

\subsection{'warm' resonance position of \ac{ECEI} channels}

\begin{figure}[htc]
   \centering
 \includegraphics[width=0.5\textwidth]{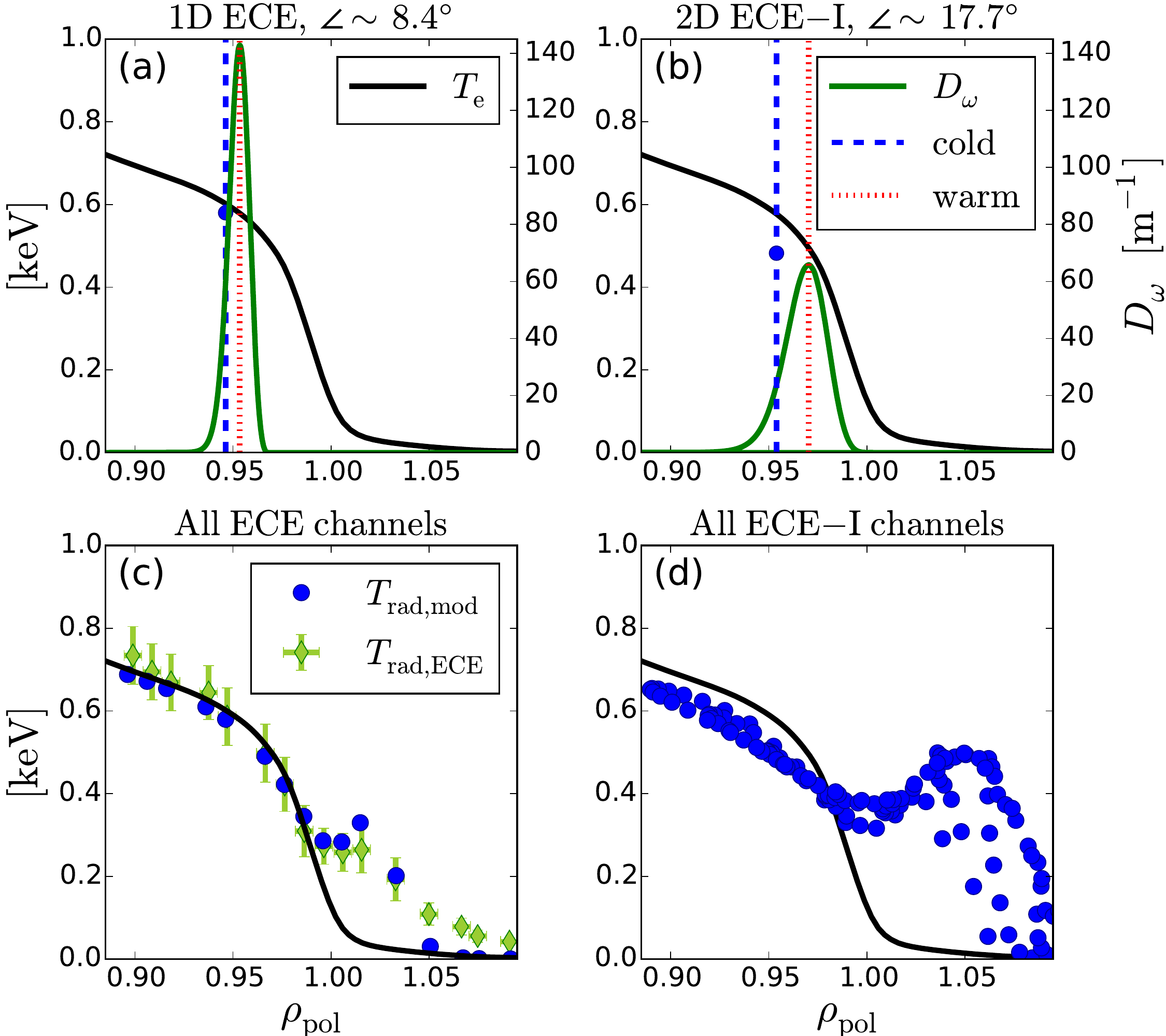} 
 \caption{Discharge $\#30839$ at $2.0\ \rm{s}$, (a) \ac{DBP} from an 1D-\ac{ECE} channel at the pedestal top is shown. Cold and 'warm' resonances are indicated by vertical dashed and dotted lines, respectively. The modeled \ac{Trad} ($T_{\rm{rad,mod}}$) is shown at its cold resonance position as blue circle. (b) same as (a) for an \ac{ECEI} channel at similar position. (c) modeled (circle) and measured  \ac{Trad} ($T_{\rm{rad,ECE}}$, diamonds) at their cold resonance position from all edge \ac{ECE} channels. (d) modeled \ac{Trad} for the \ac{ECEI} channels. The used \ac{Te} profile is plotted in all frames as a reference. }
\label{fig:1Dvs2D}
\end{figure}

Because of a toroidal inclination angle at launch of  $\angle\approx 7.2^{\circ}$ and additional poloidal  angles, the Doppler-shift influences the \ac{ECEI} even more in the case of the profile \ac{ECE} system. 
Figure \ref{fig:1Dvs2D}(a) and (b) show a comparison between the \ac{DBP} of one 1D-\ac{ECE} channel and one from the \ac{ECEI} system at similar $(R,z)$ cold resonance positions. For the shown \ac{ECEI} channel, refraction causes an angle of  $\approx17.7^{\circ}$. This leads to a significant broader \ac{DBP} and to an even larger shift between the cold and 'warm' resonance position ($\approx15\ \rm{mm}$). 

Figures \ref{fig:1Dvs2D}(c) and (d) present the modeled radiation temperature ($T_{\rm{rad,mod}}$) at their cold resonance position from all profile \ac{ECE} and \ac{ECEI} channels covering the edge, respectively.
Additionally, Fig.~\ref{fig:1Dvs2D}(c) also shows the absolutely calibrated \ac{ECE} measurements (yellowgreen diamonds). The measured and modeled \ac{Trad} are in good agreement, which underlines the correct description of these measurements. In the far \ac{SOL} ($\rho\rm{_{pol}}>1.05$), the measurements deviate from the model. These channels have a very low optical depth ($\tau_\omega<1$)~\cite{Bornatici:1983} so that wall reflections are becoming important~\cite{Denk:2016a}. But for channels having an optical depths typical for the pedestal region ($\tau_\omega>5$) or more, wall reflections contribute less than one per mill to the measured \ac{Trad}~\cite{Rathgeber:2013}. Since we are mainly interested in \ac{Trad} values from the pedestal region, wall reflections are not taken into account.

The well-known 'shine-through' peak appears in both systems~\cite{Tobias:2012} and it is even more pronounced in the \ac{ECEI} system due to their oblique \ac{LOS}. Furthermore, \ac{Trad} values differ from \ac{Te} not only in the optically thin region ($\tau_\omega<5$ for $\rho\rm{_{pol}}>1.0 $) but also in the optically thick region ($\tau_\omega>5$). This is more obvious for the \ac{ECEI} system and shows that  the classical \ac{ECE} approach~\cite{Rathgeber:2013}  is also not applicable to \ac{ECEI} channels in the optically thick region. To avoid misinterpretation of the \ac{ECEI} measurements, it is therefore required to perform the forward modeling for all channels. 

\begin{figure}[htc]
  \centering
 \includegraphics[width=0.3\textwidth]{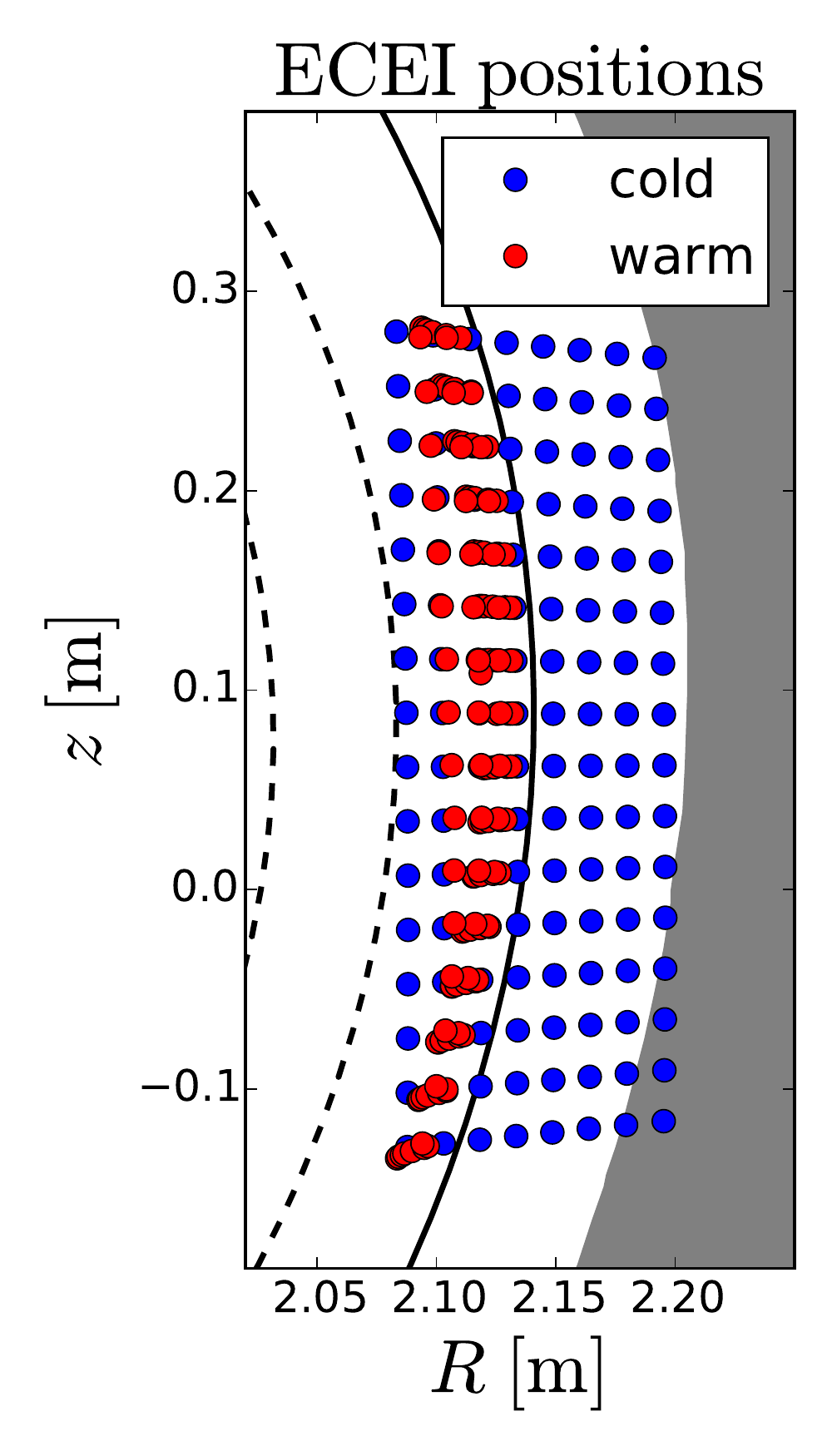} 
 \caption{ $\#30839$ at $2.0\ \rm{s}$. Cold and 'warm' resonance positions of all \ac{ECEI} channels. Differences in $R$ and $z$ between  cold and 'warm' resonances are apparent.}
\label{fig:ECEIpositions}
\end{figure}

Especially, the implementation of the ray tracing is important to determine accurate ($R, z$) values for the 'warm' resonance position.  
This is illustrated in Fig.~\ref{fig:ECEIpositions}, which shows ($R,z$) of the 'warm' and the cold resonances. Remarkably, the majority of the \ac{ECEI} channels in this configuration probe the pedestal region. 
The channels in the optically thick region measure electrons located in this region due to the Doppler-shift, whereas the \ac{SOL} channels observe only the  down-shifted radiation of the Maxwellian tail \cite{Rathgeber:2013}. One should keep in mind that the Doppler-shifted observation in the optically thick region  is a feature due to the oblique \ac{LOS}, whereas the feature of the \ac{SOL} channels probing the pedestal region appears also for the case of perpendicular \ac{ECE}  views (see Ref.~\cite{Tobias:2012}). Since the \ac{SOL} channels also contain valuable information, we will also include them in our analysis.

In the following, we will use the 'warm' resonance positions  as measurement positions and assume that they are constant during the rigid rotation.
In principle, there are two possibilities which can modify the measurement position during the rigid rotation: First, a change in the total magnetic field due the \ac{MP}-field could change  the cold resonance position and hence, the 'warm' resonance position as well. The relative strength of the external \ac{MP}-field in  front of the \ac{MP}-coils is $|\delta B| / |B| < 10^{-3}$ and around the midplane, it is even lower $|\delta B| / |B| < 10^{-4}$. The resulting difference in the resonant position using the latter case is  $\delta R < 0.2\ \rm{mm}$ at $R \approx 2\ \rm{m}$. Thus, we assume that the resonance position is constant during the rigid rotations. Second, refraction due to a 3D geometry can additionally deflect the \ac{LOS} ray and vary the measurement position.  
At the moment, the forward model of the \ac{ECE}  and the underlying ray tracing are not capable to deal with 3D flux surfaces. 
But, this additional toroidal angle is expected to be small, since $\delta \phi < \rm{arctan}(|\delta B| / |B| ) = \rm{arctan}(10^{-4} ) \approx 0.006^\circ$. Even a relatively strong $n=2$ radial flux surface displacement $\xi_r$ of $2\;\rm{cm}$ at $R \approx 2\ \rm{m}$ would result in an additional toroidal angle of only $\delta \phi <  \rm{arctan}(n\  \xi_r / R ) \approx 1.1^\circ$. 
Because of these small additional angles, they impact on the modeling of the electron cyclotron radiation transport is small, which justifies the neglect of additional refraction due to toroidal asymmetry on the ray tracing.

The combination of the forward modeling and ray tracing allows us to determine accurate ($R, z$) values of the 'warm' resonance  position for the \ac{ECE} and \ac{ECEI} diagnostics. These positions are used to calculate the \ac{SFL} angle and are, therefore, essential for its interpretation. Of course, this approach is only valid if the \ac{DBP} is not bimodal and has only one dominant peak. This is the case for the \ac{ECE} and in the majority of the \ac{ECEI} channels. Moreover, the forward modeling enables us to compare \ac{Trad} from the \ac{ECE} measurements with calculated synthetic \ac{Trad} using the 3D equilibrium from VMEC.  The generation of the synthetic data from VMEC will be elucidated in the following section.

\section{Synthetic data from ideal MHD equilibrium (VMEC)}
\label{sec:VMEC}

To compare measurements with an ideal MHD equilibrium, we use the free boundary version of VMEC~\cite{Hirshman:1986}. 
VMEC uses a variational principle to determine the shape of a set of nested flux surfaces~\cite{Hirshman:1983}.
In the free boundary version, the external \ac{MP}-field enters the calculations by the boundary condition.  The \ac{WMHD} is then self-consistently  minimized while the plasma boundary is varied to make the total pressure $\frac{1}{2 \mu_0} B^2+p = \frac{1}{2 \mu_0} B^2_V$ continuous at the plasma boundary. The normal component of the vacuum field $\vec B_V$ vanishes such as $\vec B_V \cdot \vec n_p =0$ ( $\vec n_p$ = normal vector at the plasma boundary). The vacuum field $\vec B_V$ is produced by all external conductors (e.g. toroidal field coils, shaping and perturbations coils). The converged 3D equilibrium is a nonlinear solution of the ideal \ac{MHD} model. This implies that (a) non-linear coupling of toroidal modes is correctly represented and (b) the solution preserves inherently the original topology with nested flux surfaces, i.e. magnetic islands cannot be described. This latter property corresponds to perfect shielding of resonant field components. In contrast to other formulations, the variational method ensures this intrinsically, without the need to adjust surface currents at rational flux surfaces as done e.g. in perturbative codes.

\subsection{Equilibrium inputs}
 
  The axisymmetric equilibrium reconstruction CLISTE from discharge $\#30839$ at $t=3.2\ \rm{s}$ serves as an input~\cite{McCarthy:2012}. 
  To reduce the influence of the \ac{MP}-coils on  the reconstruction, we choose a time point, when the \ac{MP}-coils located closely to the $B_\Theta$ probes have almost zero current. This is demonstrated in Fig.~\ref{fig:diags}(c), where the color scaling indicates the coil current at  $t=3.2\ s$ and the $B_\Theta$ probes are shown as well. 
  
\begin{figure}[htc]
   \centering
 \includegraphics[width=0.5\textwidth]{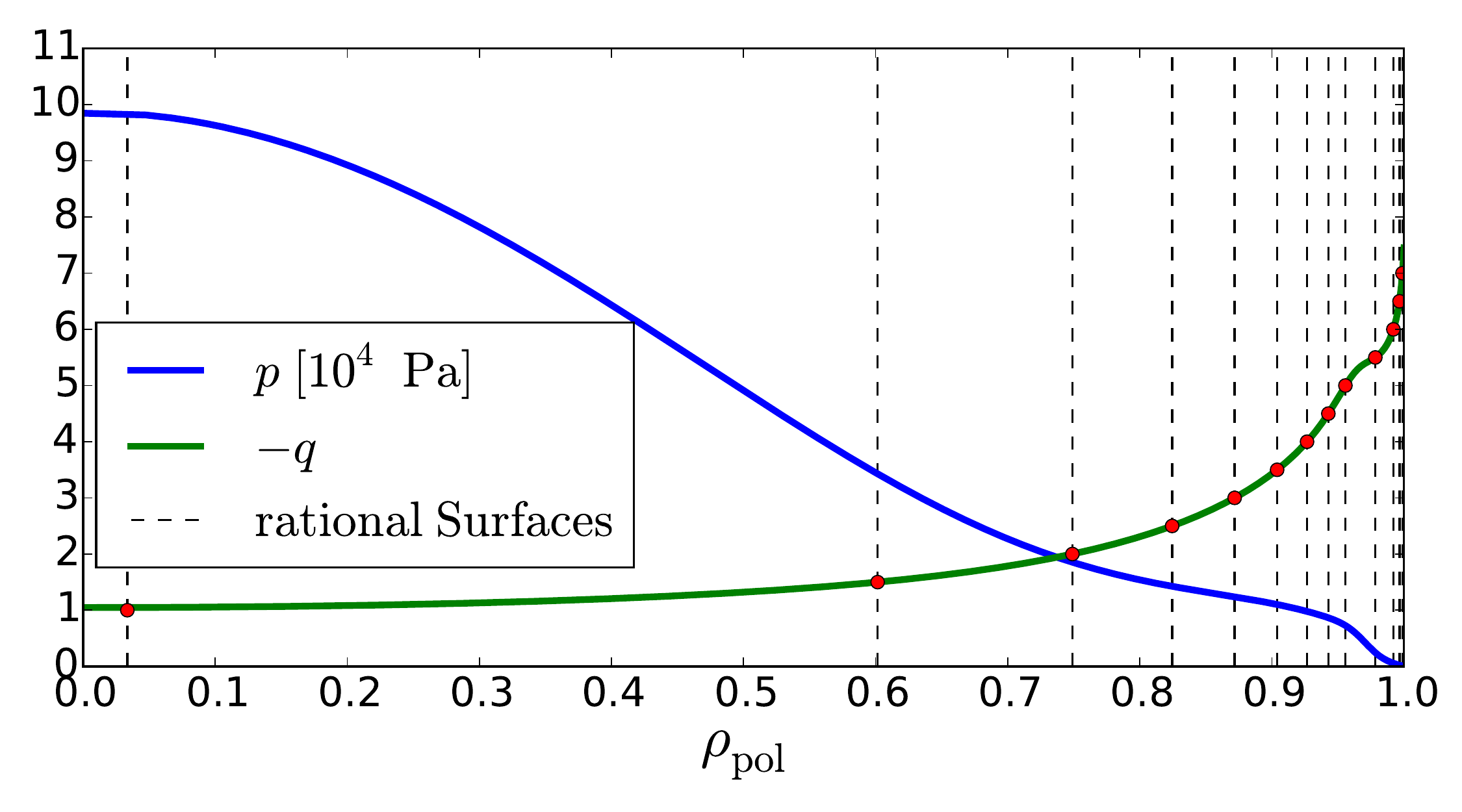} 
 \caption{Total pressure $p$ and the safety factor $q$ are shown. Vertical lines indicate the rational surface positions for $n=2$. }
\label{fig:p_q}
\end{figure}

  The pressure $p$, toroidal current and the safety factor $q$ profile (Fig.~\ref{fig:p_q}) in the CLISTE equilibrium are restricted by kinetic profiles, a self-consistent bootstrap current in the edge gradient region and a prescribed minimum $q$ above 1 to avoid an unstable helical plasma core in VMEC, respectively. For the pressure constraints, the pressure profile was determined at $t=3.2\ \rm{s}$ using  various profile diagnostics like \ac{LIB}, \ac{CXRS}, \ac{TS}, etc.   
  The contribution from the fast particles is not taken into account, but this is usually negligible in the pedestal~\cite{Dunne:2012}.
  Because of the steep density and temperature profiles in the pedestal, the resulting bootstrap current causes a flattening in the $q$ profile (around $\rho_{pol} \approx 0.97$, $q \approx 5.4$ in Fig.~\ref{fig:p_q}). Since VMEC only deals with nested flux surfaces, the  \ac{SOL} is not considered and \ac{SOL} currents were excluded in the reconstruction of the input equilibrium. Moreover, the flux surfaces were truncated at $\rho\rm{_{pol}}\approx 0.9999$ to avoid the singularity of the X-point.  One should also keep in mind that the axisymmetric solution of the converged VMEC equilibrium can slightly vary from the CLISTE equilibrium. But this difference is small and can be balanced by shifting the plasma vertically and/or radially a few millimeters ($3-4\ \rm{mm}$).
  More details about the implementation of the VMEC code at ASDEX Upgrade is described in Ref.~\cite{Strumberger:2014}. MARS-F calculations are also employed to supplement the comparison and the inputs are the same as for VMEC. 
  
 \subsection{Calculation of straight field line coordinates and poloidal mode spectra}
  \label{sec:straight}
 The calculations of the  poloidal mode spectra and the comparison to the \ac{ECEI} measurements require the usage of \ac{SFL} coordinates ($\theta^{\star}$, $\phi^{\star}$). On the flux surface $\rho$, they are defined such that $d\phi^{\star}/d\theta^{\star}=q(\rho)=const$. In this paper, we primarily use \ac{SFL} coordinates (PEST-like \cite{Johnson:1979}), where the Jacobian for the coordinate transformation with the cylindrical coordinates is proportional to $R^2$.  Moreover, the toroidal angle is the geometrical angle $\phi^{\star}=\phi$ and, thus, only the poloidal \ac{SFL} angle $\theta^{\star}$ has to be determined. For more details, we refer the readers to Chapter $6.2$ in Ref.~\cite{Dhaeseleer:1991} or Chapter 2.2.1.4 in Ref.~ \cite{Zohm:2014}. As an exception and due to historical reasons, the poloidal mode spectra of the VMEC output are calculated using Boozer coordinates~\cite{Boozer:1981}. But this makes almost no difference for the analysis since only the axisymmetric equilibrium, as in all cases,  is used to  calculate the \ac{SFL} coordinates and hence, the poloidal mode spectra.
 
\begin{figure*}
   \centering
 \includegraphics[width=1.0\textwidth]{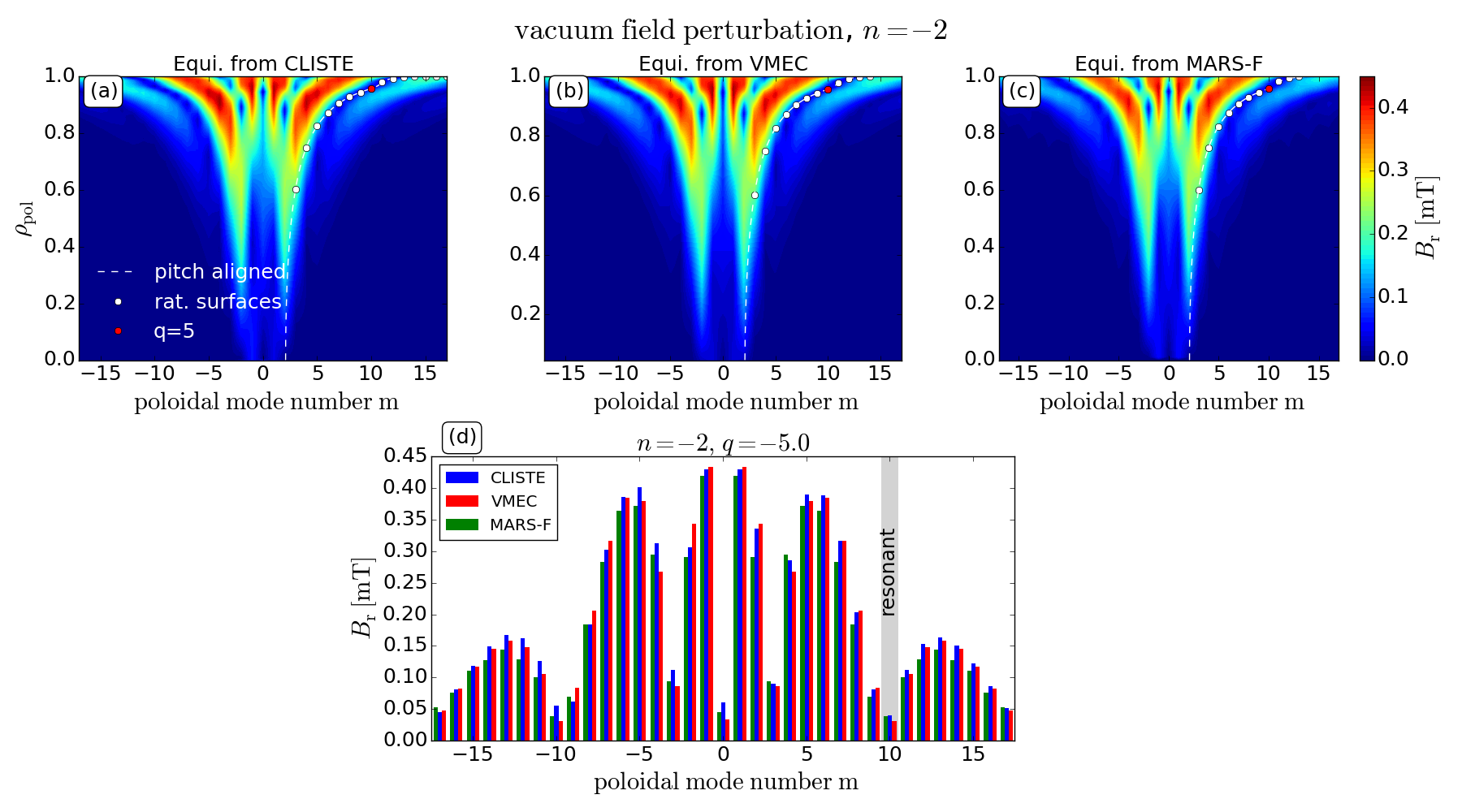} 
 \caption{The poloidal mode spectra of the external \ac{MP}-field  $n=-2$  in the vacuum field approximation, $\rho_{pol}$ versus poloidal mode number $m$
 using (a) CLISTE equilibrium of $\#30839$ at $3.2\ \rm{s}$, $n=0$ solution of (b) VMEC and (c) MARS-F. Color scaling indicates the strength of the radial field perturbation $B\rm{_{r}}$. The pitch aligned components and the rational surfaces are shown as dashed line and white circles, respectively. (d)  poloidal mode spectra of $q\approx-5.0$ ($\rho_{pol}\approx0.955$) surface indicated as red circle in (a-c). Its resonant component is illustrated by a grey bar. The external \ac{MP}-field is almost non-resonant.}
\label{fig:modespecvac}
\end{figure*}

To verify the calculations of the \ac{SFL} coordinates and the mode spectra, Fig.~\ref{fig:modespecvac} shows  the $n=-2$ poloidal mode spectra of the external \ac{MP}-field in the vacuum field approximation using the axisymmetric equilibrium from CLISTE (Fig.~\ref{fig:modespecvac}(a)), VMEC (Fig.~\ref{fig:modespecvac}(b)) and MARS-F (Fig.~\ref{fig:modespecvac}(c)). The color scaling indicates the amplitude of the perturbed magnetic field component which is normal to the unperturbed flux surface $B\rm{_{r}}$. To account for the components with the same helicity of the Fourier spectrum ($n=+2$), the illustrated $n=-2$ amplitudes are multiplied by a factor of 2. Since ASDEX Upgrade has  negative $q$ in this case, only positive poloidal mode numbers are relevant for the negative toroidal mode components. This is also seen  by the components (dashed line), which have the same helicity as the equilibrium field-line pitch (pitch aligned)  and the resonant (circle) components in Fig.~\ref{fig:modespecvac}(a-c). In general, the poloidal mode spectra are in good agreement, which gives  confidence about the representation of the external \ac{MP}-field. The mode spectra of one flux surface at the edge ($q=-5$) are illustrated in Fig.~\ref{fig:modespecvac}(d).
Slight deviations are mainly because of small differences in the axisymmetric equilibria (mm variation in position and shape) and the \ac{MP}-coil current representation (e.g. MARS-F uses Fourier representation \cite{Ryan:2015}). The analyzed coil configuration is even parity with $|n|=2$ and for the present plasma configuration with $q_{95}=-5.4$, the external \ac{MP}-field is almost non-resonant (Fig.~\ref{fig:modespecvac}(d)).  The grey bar in Fig.~\ref{fig:modespecvac}(d) indicates the resonant field component.

\subsection{Justification of using ideal MHD}
\label{sec:idealMHD}

\begin{figure}[htc]
   \centering
 \includegraphics[width=0.5\textwidth]{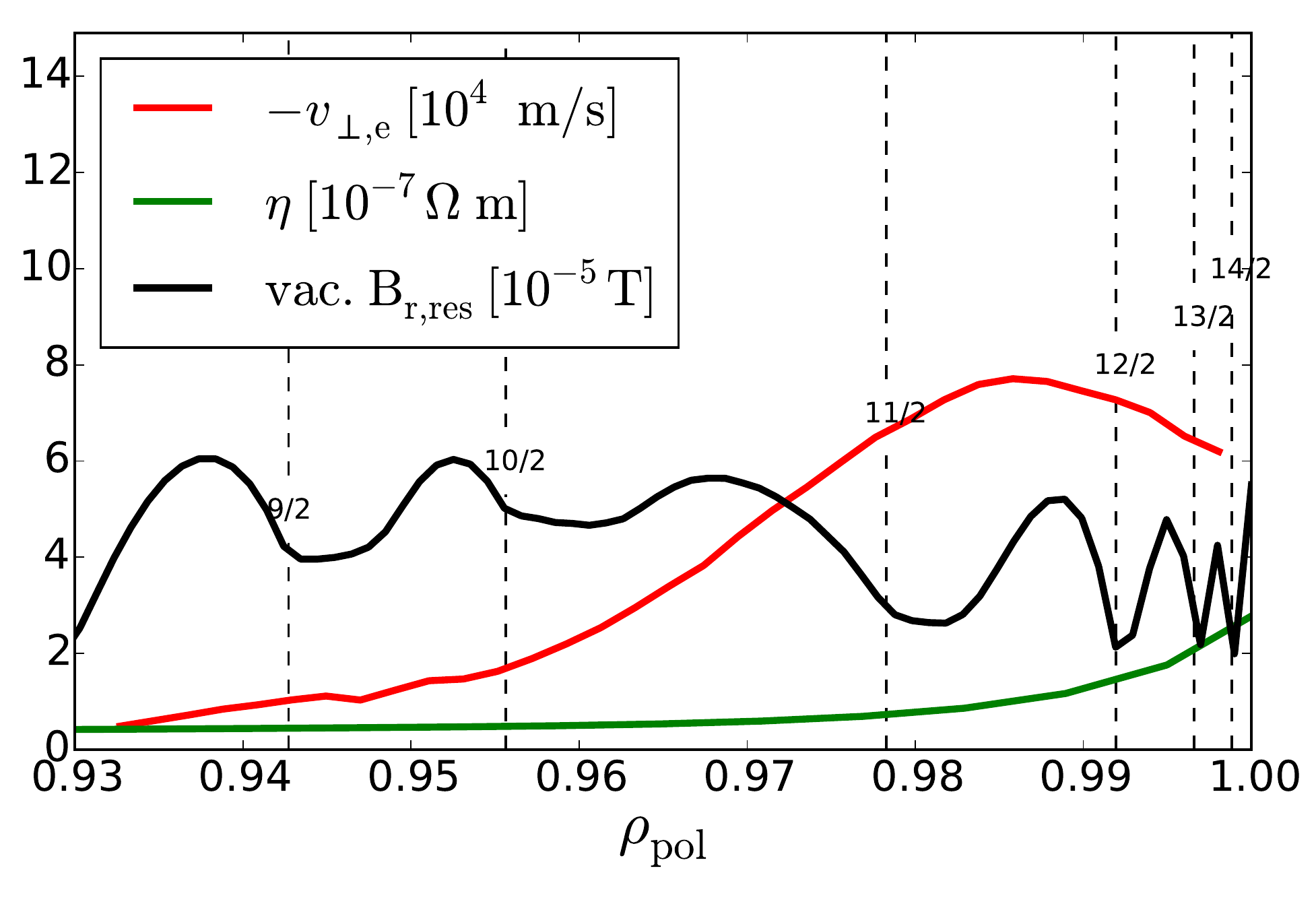} 
 \caption{ Perpendicular electron velocity $v_{\perp}$, Spitzer resistivity $\eta$ and pitch aligned radial component of the vacuum field perturbation $B\rm{_{r,res}}$ are shown. Usually high $v_{\perp}$, low $\eta$ and, for this case, low $B\rm{_{r,res}}$ are observed in the pedestal.}
\label{fig:penetration}
\end{figure} 

 Since VMEC is an ideal \ac{MHD} code with nested flux surfaces, no resistive \ac{MHD} effects are included and no magnetic island can appear. The appearance of magnetic islands at the rational surfaces  depends on the plasma resistivity $\eta$ as well as on the velocity of the plasma frame expressed by the perpendicular electron velocity $v_{\perp,e}$ and the field strength of the resonant component  of the external \ac{MP}-field  normal to the unperturbed flux surface $B\rm{_{r,res}}$. The velocity of the plasma frame in the pedestal is relatively high $v_{\perp,e}(q=-5.5)\approx -60\ \rm{km/s}$  (Fig.~\ref{fig:penetration}) due to the dominant diamagnetic velocity~\cite{Viezzer:2014}. Hence, it is expected that the, anyway small, pitch aligned components from the external \ac{MP}-field are screened. The resistivity in the pedestal region is low due to the high $T_e$. The calculated Spitzer resistivity is shown in Figure \ref{fig:penetration}. Because of the low pitch aligned components, the high $v_{e,\perp}$ and low $\eta$, magnetic islands are unlikely in this case. In order to further justify our use of ideal \ac{MHD}, we employed MARS-F calculations once with ideal MHD and once with resistive MHD using Spitzer resistivity ($\eta\rm{_{Spitzer}}$). The resulting magnetic perturbation of the plasma response field between ideal and (Spitzer) resistive \ac{MHD} differ only by maximal $0.01\ \rm{mT}$ in the poloidal mode spectra (not shown). This also implies  that the resonant components of the plasma response field are also smaller than $0.01\ \rm{mT}$.  In comparison to the values from the vacuum field calculations (see Fig.~\ref{fig:modespecvac}), this difference is very small.
 Moreover, the displacement of the resistive \ac{MHD} calculations exhibits no phase flip. This underlines the use of ideal \ac{MHD}.

\subsection{The VMEC calculation}
\label{sec:VMECcalc}

\begin{figure}[htc]
   \centering
 \includegraphics[width=0.5\textwidth]{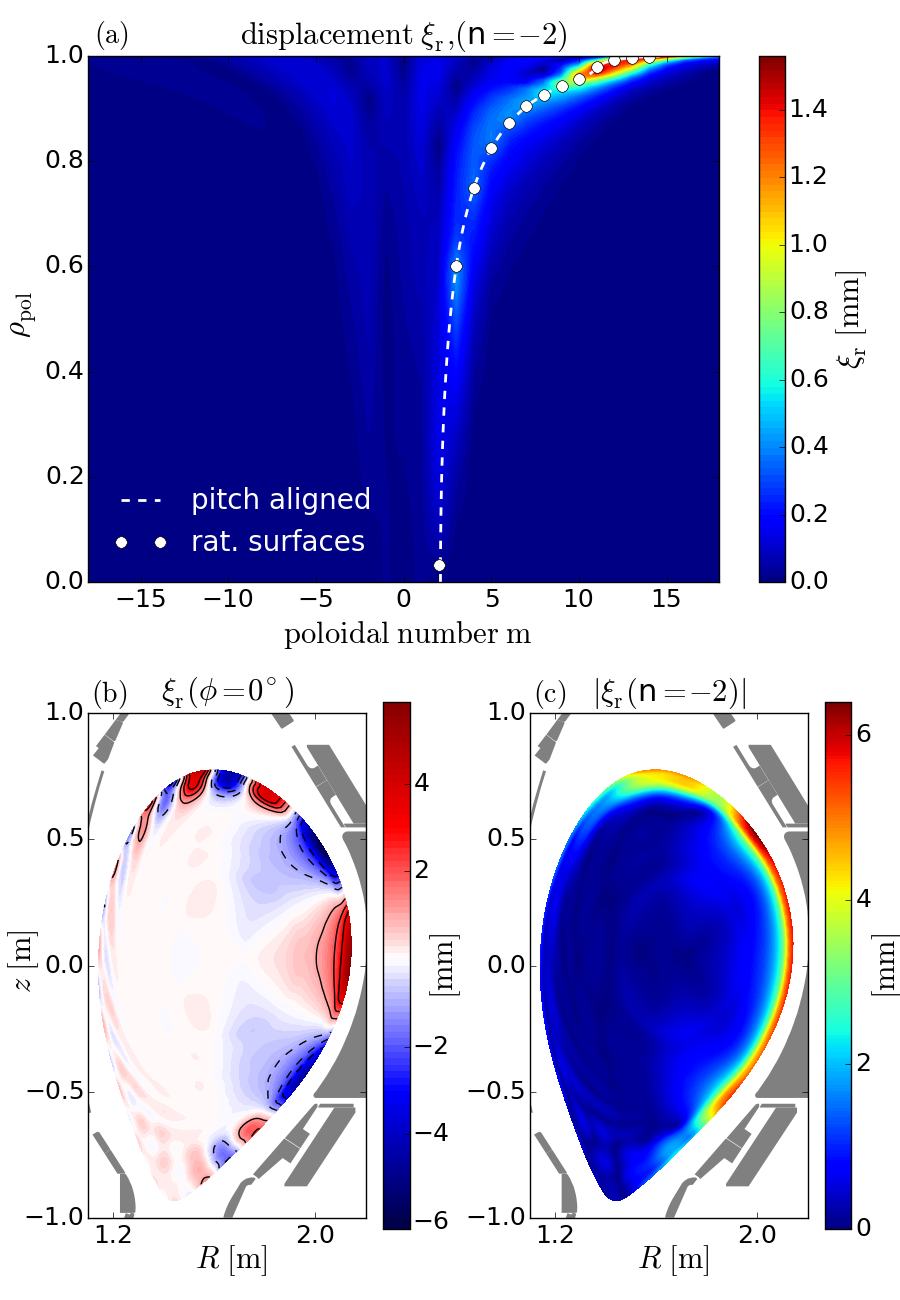} 
 \caption{ Radial displacement $\xi_r$  from VMEC, (a)  $\rho_{pol}$ versus the poloidal mode spectrum $m$ of the $n=-2$ component of $\xi_r$ amplitude, (b) poloidal cut of  $\xi_r$ at a toroidal angle of $0^{\circ}$ and (c) poloidal distribution of the $n=-2$ displacement amplitude along the toroidal coordinate. A kink response with $m\sim9-15$ at the edge ($\rho_{pol}>0.9$) around the \ac{LFS} midplane  is apparent.}
\label{fig:corrSpec}
\end{figure}

To have sufficient accuracy of the resulting equilibrium, we used 1000 flux surfaces, 17 toroidal mode numbers ($n=-8,\dots,8$) and 25 poloidal mode numbers for one period. Because of  $n=2$, only one toroidal half was calculated. The toroidal ripple was not considered.
The properties of the resulting 3D VMEC equilibrium are shown in Fig.~\ref{fig:corrSpec}. The radial displacement $\xi\rm{_r}$ is almost pitch aligned and strongest at the edge (Fig.~\ref{fig:corrSpec}(a) and details of the edge in Fig.~\ref{fig:poloidalAll}(c)). The poloidal mode spectra can also be seen from a poloidal cut $\phi=0^\circ$ of $\xi\rm{_r}$ (Fig.~\ref{fig:corrSpec}(b)).  The amplitude of the $n=-2$ displacement along the toroidal coordinate is shown in Fig.~\ref{fig:corrSpec}(c). It  is  pronounced around the midplane at the \ac{LFS} (Fig.~\ref{fig:corrSpec}(b)).

The VMEC calculations also exhibit a small $n=4$ component, which can solely be attributed to the plasma response (not shown). The explanation is as follows: ASDEX Upgrade has 8 saddle coils in each row. Hence, the  $n=2$ perturbation can be described as a rectangular function along the geometrical toroidal angle $\phi_{geo}$. The Fourier series of a rectangular function solely consists of odd harmonics ($1,3,5,\dots$). 
Consequently,  the applied $n = 2$ \ac{MP}-field  in the vacuum approximation has exclusively  toroidal mode numbers of $n=2,6,10,\dots $. Additionally, the $n=6$ component is increased due to the aliasing effect from the $n=2$ perturbation using 8 saddle coils ($n\rm{_{aliasing}}=8 - 2$).
However, ASDEX Upgrade has no $n=4$ component in the vacuum field spectra when applying an $n=2$ perturbation. In DIII-D, for example, this is not the case. It has 6 saddle coils in each row and the aliasing effect causes an $n=4$  component~\cite{Haskey:2014}. 
The combination of Ampere's law $\vec{j}=\vec{\nabla} \times \vec{B}$ and the plasma equilibrium $\nabla \vec{p}=\vec{j} \times \vec{B}$ introduces a non-linear ($\approx B^2$) behavior due to the plasma response. This non-linearity can lead to the appearance of  additional toroidal mode numbers like $n=4,8,12,\dots$. Therefore, a measured $n=4$ component would prove a plasma response. But according to VMEC the maximum displacement of $n=4$ amounts to $\xi_{r}(n=4) \approx 0.2\ \rm{mm}$ and is unlikely to be measured within the measurement accuracy.


\subsection{Calculation of synthetic  data}
\label{sec:syntheticECE}

The output of VMEC is a 3D equilibrium calculated for one time point. To compare the toroidally localized measurements during a rigid rotation with this single 3D equilibrium, we developed synthetic diagnostics for the VMEC equilibrium. The most important steps to produce synthetic data are listed below:
\begin{itemize}
      \item [(i)] The currents of the \ac{MP}-coils are used to map the timebase of the used diagnostic  to the geometrical toroidal angle $\phi\rm{_{geo}}$ in VMEC or vice versa. It is mapped in such way that the calculated $\phi\rm{_{geo}}$ from the rotation corresponds to the toroidal position of the diagnostic at the time of the VMEC calculation ($t=3.2\ \rm{s}$). Each slice in $\phi\rm{_{geo}}$ can be correlated to a time point and vice-versa.
      \item[(ii)] Because of small discrepancies between the input equilibrium and the axisymmetric component of the VMEC solution, the entire VMEC equilibrium is shifted by $R=3\ \rm{mm},\ z=4.5\ \rm{mm}$ to align the  surfaces at the \ac{LFS}. This allows us to compare vacuum field calculations using the input equilibrium with VMEC at the \ac{LFS}.
      \item[(iii)] The (R, z) positions of each channel are determined for each diagnostic and are assumed to be independent of time or rather $\phi\rm{_{geo}}$. In the case of \ac{ECE} diagnostics, the 'warm' resonance positions are used.
    \item[(iv)] $T_e$, $T_i$ and $n_e$ profiles before the \ac{MP} onset are used to correlate the $\rho\rm{_{pol}}$ with $T_e$, $T_i$ and $n_e$ values assuming they are constant on the perturbed flux surface. Due to a slight increase in core $n_e$, the global $T_e$ slightly decreases. To account for this, we add a time or rather $\phi\rm{_{geo}}$ dependent scaling function. This function is a cubic spline and  time traces from core channels were used to parametrize it.
   \item[(v)] ($R_i$, $z_i$) of each channel $i$ and $\phi\rm{_{geo}}$ are used to deduce the corresponding  $\rho\rm{_{pol}^i}(\phi\rm{_{geo}})$  values from the 3D VMEC equilibrium and therefore, also $T_i$ and $n_e$  values. To get synthetic \ac{Trad} values for \ac{ECE} diagnostics, the electron cyclotron radiation transport is solved using the  slices of the  poloidal flux surface at the corresponding $\phi\rm{_{geo}}$ of the perturbed equilibrium. To solve radiation transport, $T_e$, $n_e$ profiles as in step (iv) are used and each slice in $\phi\rm{_{geo}}$ is assumed to be axisymmetric.
    \item[(vi)] The VMEC output has no \ac{SOL} flux surfaces. To complete the synthetic profiles in the \ac{SOL}, the CLISTE equilibrium is used for flux surfaces for $\rho\rm{_{pol}}>1.04$. To allow a smooth transition between the perturbed VMEC and the axisymmetric CLISTE equilibrium, we simply use a 2D cubic spline to interpolate in-between.  
 \end{itemize}

All these steps allow us to compare quantitatively synthetic data from VMEC with measurements from \ac{ECEI}, \ac{ECE}, \ac{CXRS} and \ac{LIB}. Moreover, we distinguish between the comparison of the amplitude and phase or rather poloidal mode structure of the flux surface displacement.
We deduce both by fitting the measured and synthetic data to sine function with its harmonics. 

\section{Amplitude comparison}
\label{sec:amplitude}


 \begin{figure}[htc]
   \centering
 \includegraphics[width=0.5\textwidth]{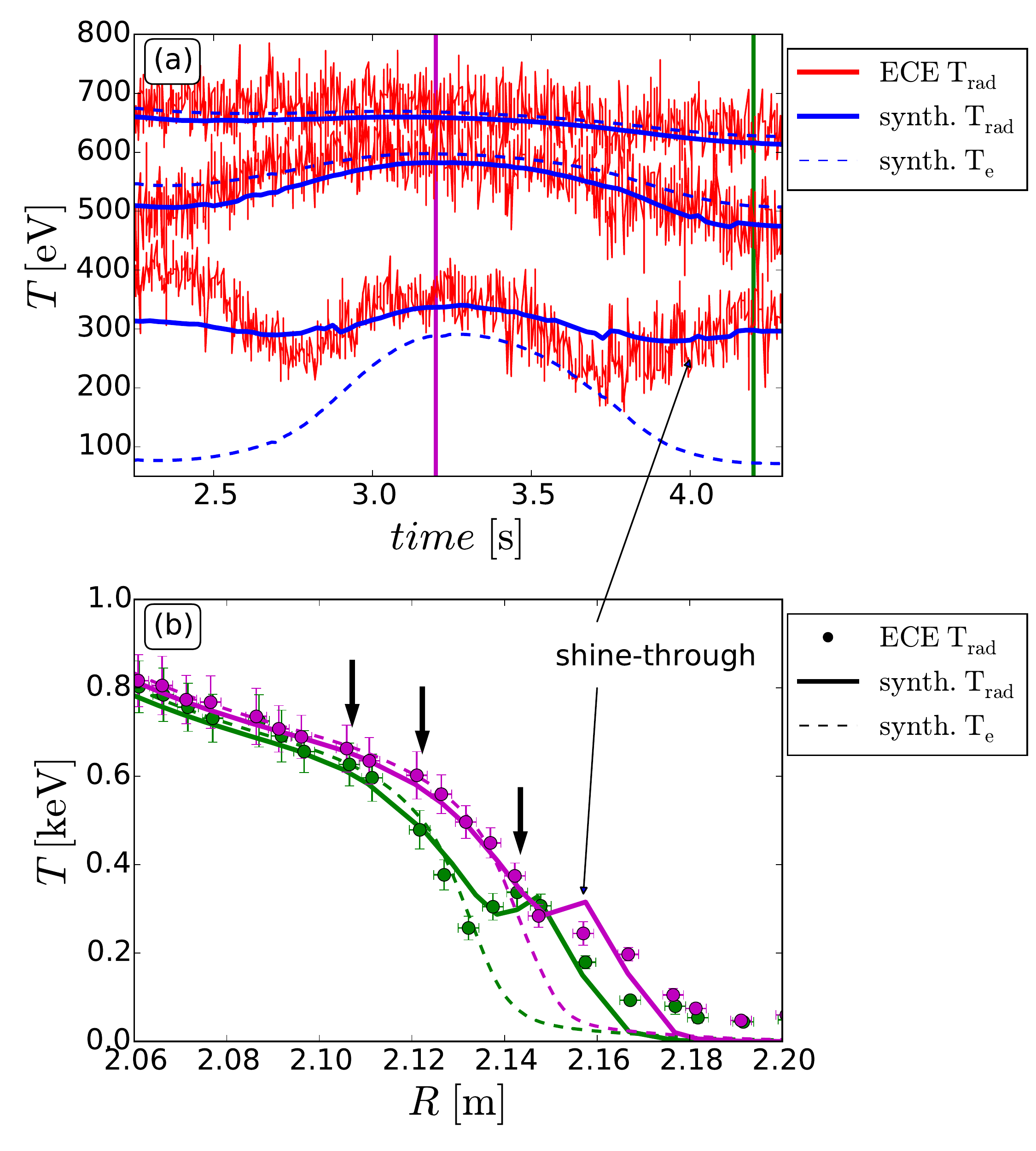} 
 \caption{$\#30839$, (a) time traces of three channels using synthetic \ac{Te} (dashed blue), \ac{Trad} (solid blue), and the \ac{ECE} data  (solid red). Their radial positions are indicated by arrows in (b). (b) Synthetic \ac{Te} (dashed), \ac{Trad} (solid) and measured \ac{Trad} profiles (circles) at maximum (magenta) and minimum (green) displacement. The corresponding time points are indicated by the colored vertical bars in (a). Deviations between synthetic \ac{Te} and \ac{Trad} due to the 'shine-through' effect are indicated. The synthetic and measured  \ac{Trad} profiles are in good agreement.}
\label{fig:ECEsynProf}
\end{figure} 

In the following, we focus on the profile \ac{ECE} system. Unlike the \ac{ECEI}, it has an absolute calibration and moreover, its spatial resolution is higher. Both is beneficial for the amplitude comparison.
But like every \ac{ECE} system, its interpretation at the plasma edge can be challenging due to the transition from the optically thick to the optically thin plasmas (also discussed in Ref.~\cite{Tobias:2013a}). Because of the lack of density information  at the \ac{ECE} \ac{LOS}  in the  presence of a 3D equilibrium perturbations, we are not able to accurately determine \ac{Te} profiles from \ac{ECE} measurements. Therefore, we will primarily compare \ac{Trad} data instead of \ac{Te}. 

Figure \ref{fig:ECEsynProf} shows a comparison between \ac{ECE} measurements, synthetic \ac{Trad} as well as synthetic \ac{Te} data. Both, the time traces (Fig.~\ref{fig:ECEsynProf}(a)) and the corresponding profiles (Fig.~\ref{fig:ECEsynProf}(b)) from  the \ac{ECE} and the synthetic \ac{Trad} diagnostic match very well. The boundary displacement is visible as a shift between between the profiles at the minimum and maximum displacement (magenta versus green in Fig.~\ref{fig:ECEsynProf}(b)).  Moreover, the synthetic \ac{Trad} profiles correctly describe features of the 'shine-through' peak. For example, the gradients of the edge \ac{Te} profile are smaller at the maximum displacement (magenta profiles in Fig.~\ref{fig:ECEsynProf}(b)) and therefore, the 'shine-through' peak is less pronounced (lower in height and broader) with respect to the \ac{Te} profile. This is seen in the measurements as well as in the synthetic data, but less distinct. This also indicates that VMEC slightly underestimates the change in the edge gradients. Moreover, the displacement seems also to be underestimated.
As already mentioned, channels in the far \ac{SOL} measure a higher radiation than expected from the model. This is because the model does not take wall reflections into account, which are particularly important at very low optical depth.

Since we are dealing with multichannel profile diagnostics, it is possible to compare the displacement using either the entire profile information or it is also possible to use the amplitude information from the individual channels. Both possibilities will be discussed in the subsequent sections.

\subsection{Using single channel information from ECE}

\begin{figure}[htc]
   \centering
 \includegraphics[width=0.5\textwidth]{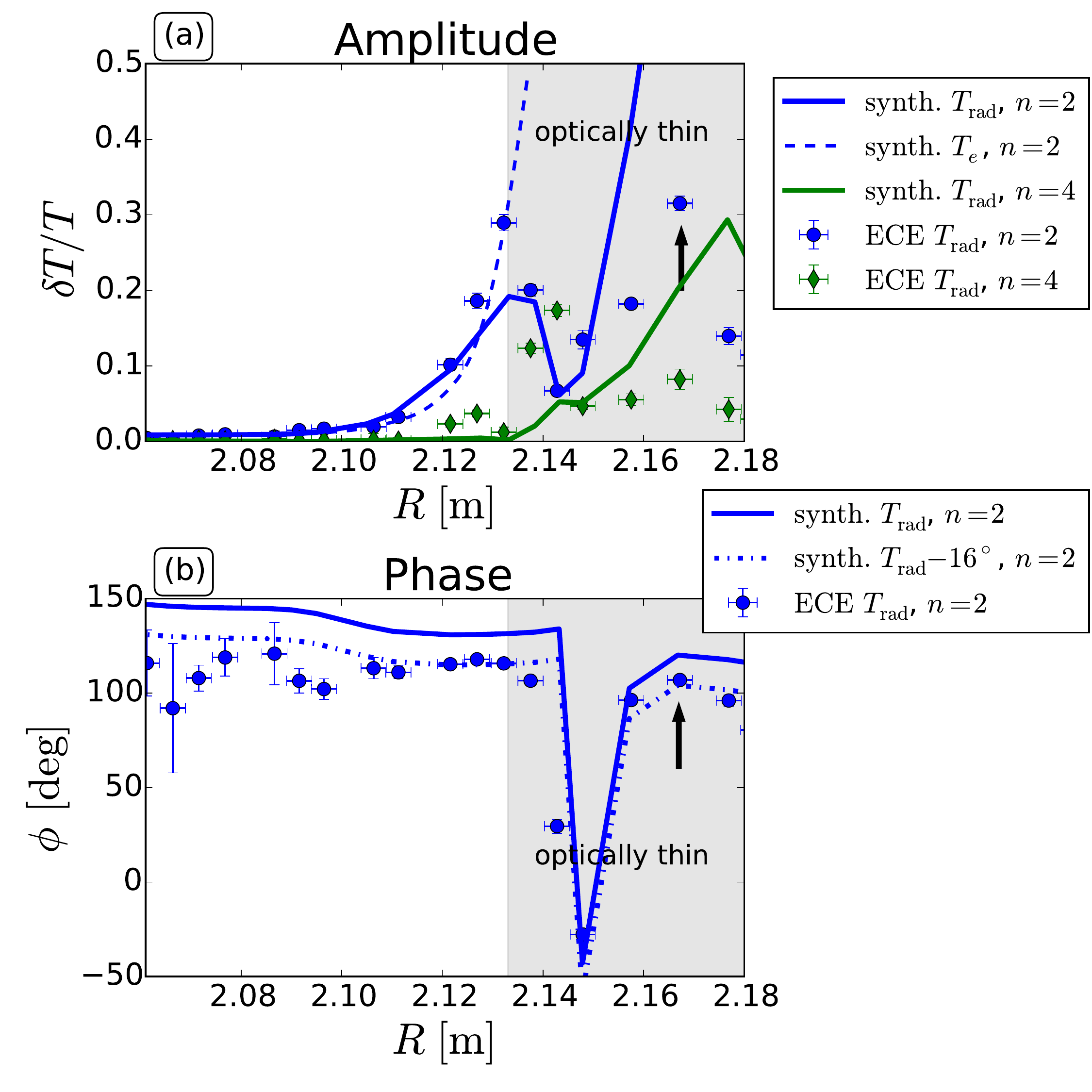} 
 \caption{$\#30839$, (a) relative amplitude of the $n = 2$ (blue) and $n = 4$ (green) component from synthetic \ac{Trad} (solid), and \ac{ECE} measurements (circle) as well as only $n = 2$ from synthetic \ac{Te} (dashed). (b) phase of  $n = 2$ from synthetic \ac{Trad} and  \ac{ECE} . The phases systematically deviate by around $16^{\circ}$ indicated by the dash-dotted line. Channels in the optically thin region can significantly distort the amplitude and phase analysis.}  
\label{fig:ECEAmp}
\end{figure}

This analysis is based on fitting each ECE channel using a sine function including its harmonics to extract the amplitude information.  The 'shine-through' effect limits the usage of \ac{Trad} to investigate edge perturbations. 
Moreover, it can lead to a misinterpretation of the amplitudes and phases from an $n=2$ perturbation and can lead to a misapprehension of an '$n=4$' component. 
This is illustrated in Fig.~\ref{fig:ECEAmp}, where the amplitude and the phase of the synthetic \ac{Trad}, \ac{Te} and the \ac{ECE} measurements (ECE \ac{Trad}) are shown. Figure \ref{fig:ECEAmp}(a) shows the relative amplitudes ($\delta T/T$) of $n=2$ and $n=4$. The $n=2$ amplitude of the synthetic \ac{Te} (blue dashed) decreases  from  the edge towards the plasma core. 
The 'shine-through' effect corrupts the simple analysis of phase and amplitude using \ac{Trad} from single channels. Especially the channels, which are located in the 'shine-through' well, are affected (channels around $R\approx2.15\ \rm{m}$ in Fig.~\ref{fig:ECEAmp}). These channels view alternating the optically thin and thick region throughout the rotation. As a result, they show a reduced $n=2$, an increased $n=4$ amplitude and a distorted phase.  This is clearly seen in the \ac{ECE} measurements and well captured by the synthetic \ac{Trad}  data.
This measured '$n=4$' component is most likely an artifact from the 'shine-through', because the $n=4$ component, according to VMEC, amounts only to about $1/30$ of the $n=2$ component. 
 On the positive side, this '$n=4$' amplitude and the distorted phase can be used to exclude the corrupted channels without knowing its exact measurement position. We use this simple recipe to exclude  \ac{ECEI} channels viewing the 'shine-through' well.
In the far \ac{SOL},  discrepancies between measurements and synthetic \ac{Trad} are apparent. This is because these channels still observe some radiation due to wall reflections.

The amplitude as well as its decay of \ac{ECE} and synthetic \ac{Trad} data in the optically thick region are in good agreement. The amplitude is mainly measurable at the edge. This is because the perturbation is localized at the edge and measurements of the displacement are more sensitive  in the large gradient region. Hence, the measured amplitude is a convolution between an edge perturbation and a localized \ac{Te} gradient. This makes a quantitative comparison using single channel information difficult, because its correct interpretation depends highly on the correct resonance positions of the \ac{ECE} channels.

In contrast to the amplitude comparison, the phase information is less dependent on the radial position. The boundary perturbation around the \ac{LFS} midplane has a relatively large poloidal extension  and  the distortion penetrates straightly from the edge towards the core (see Fig.~\ref{fig:corrSpec}(b)). Thus, the phase is not changing much along the  \ac{ECE} \ac{LOS} within the optically thick region ($R<2.13\ \rm{m}$). This is seen in the \ac{ECE} measurements and confirmed by the synthetic \ac{Trad} data (Fig.~ \ref{fig:ECEAmp}(b)). Moreover, the synthetic \ac{Trad}  successfully describes the phase flip at the edge ($R\approx2.14-2.15\ \rm{m}$), which is caused by the interplay between the displacement and the non-monotonic characteristics of the \ac{Trad} profile. Channels viewing only the optically thin plasma also contain the phase information from the pedestal. The ECE channels in the \ac{SOL}, that  show a strong $n=2$ component (one is indicated by an arrow in Fig.~\ref{fig:ECEAmp}), observe almost the same phase as the pedestal channels ($R\approx2.12\ \rm{m}$). This is expected because they measure the down-shifted radiation of the electrons in the pedestal region. Slight differences in the phase between these channels in the optically thick and thin region (e.g.~arrow in Fig.~\ref{fig:ECEAmp}) can occur, because their ray paths differ slightly as well. The reason for this lies in the different measurement frequencies (see Sec.~\ref{edgeMeasurements}), which can have an impact on the ray refraction.


The phase profile of  \ac{ECE} \ac{Trad} and synthetic \ac{Trad}  shows a systematic offset of $\Delta \phi \approx 16^{\circ}$ between them (indicated by the dash-dotted line in Fig.~\ref{fig:ECEAmp}(b)). This offset can be explained by the \ac{PSL} response. The \ac{MP}-coils are mounted close to the \ac{PSL}. Since it is a copper conductor, image currents in the \ac{PSL} can  screen transient magnetic fields produced by the \ac{MP}-coils. Thus, the \ac{PSL} acts similar to $L/R$ lowpass filter and delays the rigid rotation~\cite{Suttrop:2009b}. Although the \ac{MP}-field was  rotated  by only $0.5\;\rm{Hz}$, the \ac{PSL} causes a measurable phase delay with respect to the applied coil current.  This is also inline with newly employed finite elements calculations of the \ac{MP}-coils and the \ac{PSL}, which predict a phase delay in the midplane of $\Delta \phi\rm{_{upper}} \approx 14.3^{\circ}$ regarding the upper coil set and $\Delta \phi\rm{_{lower}} \approx 11.2^{\circ}$ for the lower one~\cite{Suttrop:2009a, Suttrop:2016}. The phase delay of the upper and lower coil set is different because of the different geometry with respect to the \ac{PSL}.

\subsection{Comparison using profile diagnostics}

\begin{figure*}[htc]
   \centering
 \includegraphics[width=1.\textwidth]{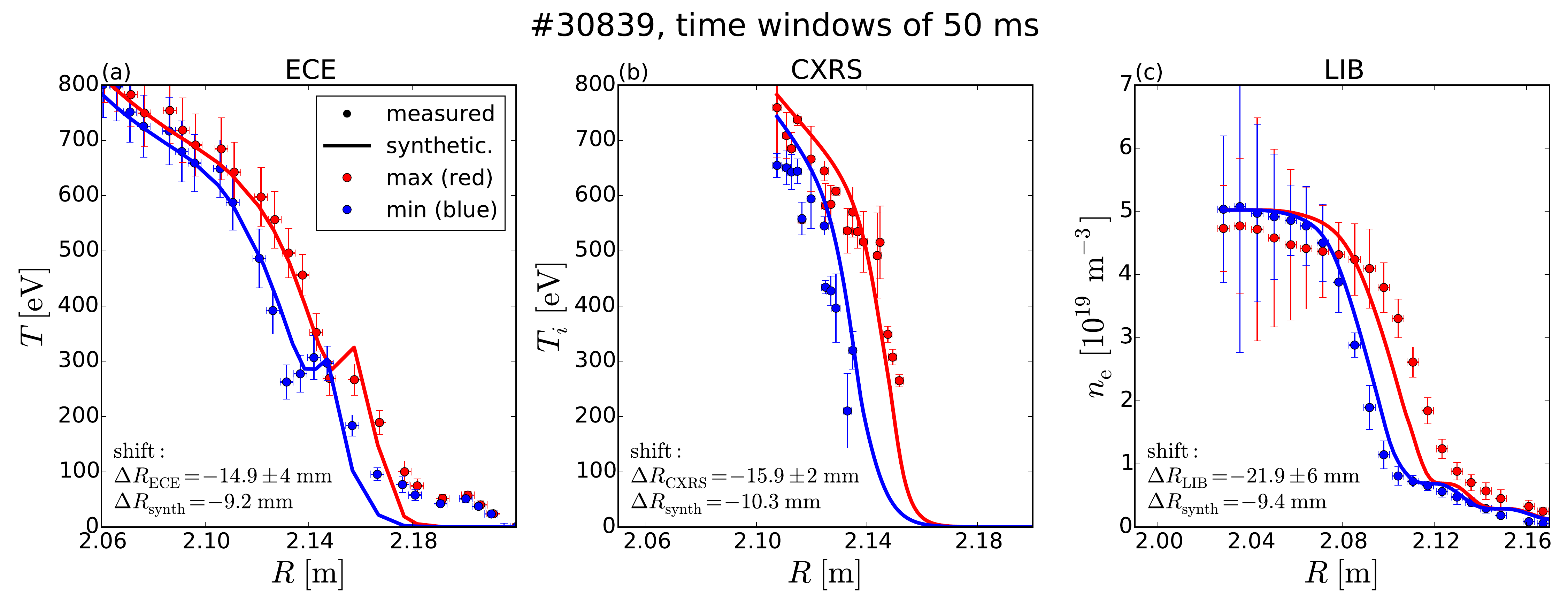} 
 \caption{$\#30839$, profiles from (a) \ac{ECE} (b) \ac{CXRS} and (c) \ac{LIB}. Measured (circles) and synthetic (solid) profiles at the maximum (red) and minimum displacement (blue) of each diagnostic. The analyzed time windows are $50\ \rm{ms}$. The derived displacements are given in the left bottom corner. \ac{ECE} and \ac{CXRS} observe the same displacement, whereas the one measured by \ac{LIB} is slightly larger.}
\label{fig:allAmp}
\end{figure*}

In the previous section, we used the amplitude information from single channels to compare it with synthetic data. But it is also possible to use the information from the entire edge profile. The displacement can be directly obtained by the radial shift between two profiles at the time of the maximum and the minimum displacement. Moreover, it also allows us to compare the displacement between the different profile diagnostics even if the measured plasma parameters are not the same. Figure \ref{fig:allAmp} shows \ac{ECE}, \ac{CXRS} and \ac{LIB} profiles at the time of maximum and minimum displacement. The corresponding synthetic data are also plotted. Due to the different toroidal and poloidal arrangement of the diagnostics, the times of the maximum and minimum vary for the different diagnostics. In general, the agreement between the synthetic and the measured profiles is good. To get one quantity for the displacement, first, we fit the profiles at the maximum displacement using a spline. Then, this spline is only varied by a radial shift until the \ac{LSQ} is minimized using the data at the minimum displacement. This is relatively robust and the uncertainties due to the change in the gradients are also reflected in the uncertainties of the determined shift. 
Because of the 'shine-through' and dominating passive lines, the \ac{ECE} and the \ac{CXRS} data from the \ac{SOL} are not used for this procedure. The same procedure was also applied to the synthetic data generated from the VMEC equilibrium. The displacements derived from this method are given in the left bottom corner of each frame in Fig.~\ref{fig:allAmp}. \ac{ECE} and \ac{CXRS} deliver very similar displacements, but they exhibit slightly larger values than predicted by VMEC. In the case of \ac{LIB}, this difference between the measured and the synthetic data is more pronounced.

\begin{figure}[htc]
   \centering
 \includegraphics[width=0.5\textwidth]{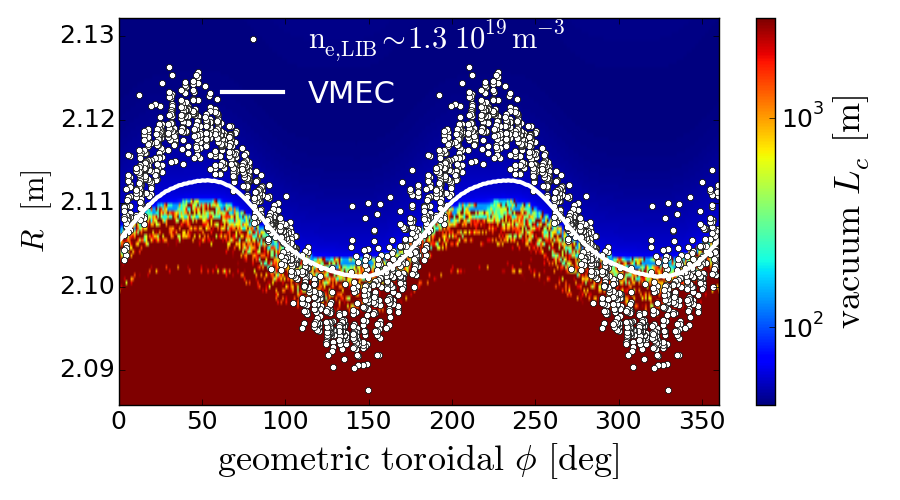} 
 \caption{$\#30839$, circles are the estimated separatrix from \ac{LIB} using a density of $1.3\cdot 10^{19}\ \rm{m^{-3}}$. Solid line is the outermost flux surface from VMEC along the \ac{LIB} \ac{LOS} and the color scaling indicates the connection length $L_C$ using the vacuum field approximation. \ac{LIB} data exceeds the VMEC and the vacuum field calculations. \ac{LIB} data is shifted radially inwards by $2\ \rm{mm}$.}
\label{fig:LIBAmp}
\end{figure}

Unlike \ac{ECE} and \ac{CXRS}, the \ac{LIB} diagnostic is well-suited for determining changes of the separatrix position. Assuming a constant separatrix density during the rigid rotation, the separatrix position can be easily tracked along the \ac{LIB}. We determine the separatrix density using the density profile prior to the \ac{MP} onset ($n_{e,sep}\approx1.3\cdot 10^{19}\ \rm{m^{-3}}$ from Fig.~\ref{fig:LIBprofiles}(a)).
Figure~\ref{fig:LIBAmp} illustrates (i) the separatrix position determined from the \ac{LIB} diagnostic , (ii) the outermost boundary of the VMEC equilibrium and (iii) the boundary from the vacuum field approximation is indicated by the color scaling using the connection length $L_c$ of the stable manifold (see Ref.~\cite{Moyer:2012}). Both, (ii) and (iii) are calculated along the \ac{LIB}.
The displacement from the VMEC equilibrium exceeds the prediction from vacuum field calculations by a few millimeters. The sinusoidal is well seen in  the measurements and agrees qualitatively. This comparison indicates a larger displacement than predicted by VMEC and is also consistent with the measurements shown in Fig.~\ref{fig:LIBAmp}. 

\subsection{Discussion of the amplitude comparison}

A quantitative amplitude comparison using single \ac{ECE} channels is challenging due to the dependencies on the measurement positions and the 'shine-through' effect. Small variations in the position can have large influence on its interpretation. Despite these difficulties, the decay of the distortion towards the plasma core agrees very well between the \ac{ECE} \ac{Trad} and the synthetic \ac{Trad}. 
The displacement can be analyzed using  the relative amplitude from single channels. This method makes a comparison between different diagnostics difficult, because they measure different plasma parameters. A comparison to e.g.~VMEC requires the development of synthetic diagnostics.
Thus, it is more useful to determine the displacement by aligning the entire edge profiles. This allows us to compare directly and quantitively the measured displacement not only to others diagnostics, but also to 3D equilibrium codes. 

All edge profile diagnostics around the \ac{LFS} midplane exhibit a displacement, which is slightly larger than predicted by the VMEC equilibrium and thus, larger than calculated in the vacuum field approximation as well (see Fig.~\ref{fig:LIBAmp}). 
In principle, the plasma position control could artificially amplify or mitigate the distortion. This depends on the relative phase between the position of the  $B_{\Theta}$ arrays and the used profile diagnostics (discussed in Ref.~\cite{Chapman:2014b}). We can exclude this in the midplane for two reasons. First, \ac{CXRS} and 1D-\ac{ECE} measure the same amplitude at different  toroidal phases (see time traces in Fig.~\ref{fig:edgetraces}). Second, the \ac{CXRS} system is in the midplane on the opposite side of the $B_{\Theta}$ array (see Fig.~\ref{fig:diags}).
A feedback controlled system solely based on measurements of one toroidal position would counteract the 3D distortion~\cite{Chapman:2014b} and, therefore, the modulation at the position of the \ac{CXRS} system would be compensated. Because of the fact that the displacement from edge \ac{CXRS} also exceeds the prediction from VMEC (see Fig.~\ref{fig:allAmp}(b)), we assume that the plasma position control system does not artificially amplify the measured modulation in the midplane. The position control system of ASDEX Upgrade also uses toroidal flux loops for the feedback control system, which seem to mitigate the effect of the \ac{MP}-field on the control system in comparison to other devices like MAST~\cite{Chapman:2014b}. However, small changes in the shape due to the control system can certainly not be excluded, which could explain that  \ac{LIB} measures a larger displacement than \ac{CXRS} and \ac{ECE}. 

VMEC seems to slightly underestimate the displacement in the midplane. A quantitive comparison of MARS-F employing the resistive as well as the ideal \ac{MHD} model with VMEC using the identical inputs show very similar displacement values. This indicates that the used input parameters can also be responsible for this underestimation. As already mentioned in section \ref{sec:VMECcalc}, the used pressure profile was determined by aligning various diagnostics at one time point during the \ac{MP}-phase. The resulting total pressure has experimental uncertainties because the used profile diagnostics are toroidally separated. As shown in Fig.~\ref{fig:LIBprofiles}, the gradients can vary depending on the toroidal phase. Since the amplitude of the displacement or rather the stable ideal kink modes are  driven by the edge pressure gradient and the associated bootstrap current, a lower input pressure gradient can lead to a smaller displacement in \ac{MHD} equilibrium codes.  This sensitivity should be kept in mind. For further details on the sensitivity  of the displacement on the pressure profile and hence, the plasma beta, we refer to the sensitivity studies in Ref.~\cite{Strumberger:2014,Paz-Soldan:2016}.


\section{Poloidal mode structure comparison}
\label{sec:poloidal}
 
 The measured displacement and the one from VMEC exceed the prediction from the vacuum field calculations. This indicates the presence of a kink response, which amplifies the magnetic perturbations and thus, the displacement. According to plasma response calculations, these amplified magnetic perturbations show dominant non-resonant components ($|m|>|nq|$). To investigate if this non-resonant behavior is also seen in the  \ac{ECEI} system, we make use of its poloidal resolution and compare the measured data to VMEC calculations.

\subsection{\ac{ECEI} vis-a-vis VMEC}

\begin{figure}[htc]
   \centering
 \includegraphics[width=0.5\textwidth]{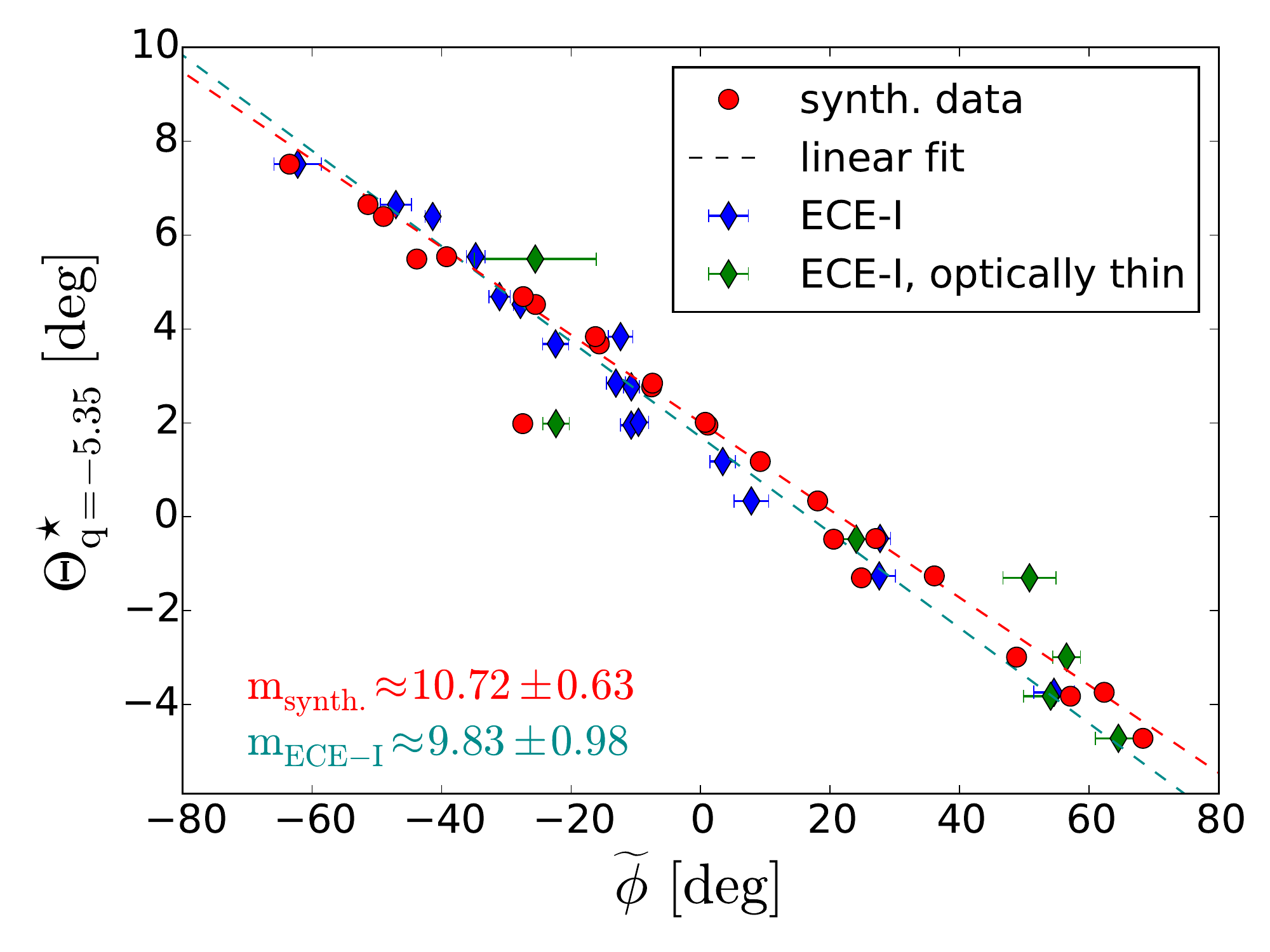} 
 \caption{$\#30839$, \ac{SFL} angle using the $q=-5.35$ flux surface versus the derived phase delay using $\widetilde{\phi} = \phi \ - <\phi>$. ECEI data from optically thick (blue diamonds) and thin (green diamonds)  plasma regions  agree with synthetic data from VMEC  (red circles). The poloidal mode numbers $m$ (left bottom corner) are determined from the slope of the linear fits (dashed).}
\label{fig:ECEphase}
\end{figure}

Figure \ref{fig:ECEphase} shows the comparison between the measured (diamonds) and the synthetic \ac{ECEI} (circles) data. To avoid corruption from the 'shine-through' well, channels are discarded which measure a significant '$n=4$' (${\delta T\rm{_{rad}}}/{T\rm{_{rad}}}(n=4)>0.05$) component. We only use channels with a significant $n=2$ component (${\delta T\rm{_{rad}}}/{T\rm{_{rad}}}(n=2)>0.085$). 
The few channels, which have permanently their cold resonance position  in the optically thin region and fulfill the mentioned conditions, are also taken into account (green diamonds). 

All selected channels from \ac{ECEI} and their corresponding synthetic channels are fitted using the \ac{LSQ} fit of a sine function including higher harmonics. The $\rho_{pol}$ values of the used channels range from $0.95$ to $0.981$. Their mean value is $0.968$, which is the  $q\approx-5.35$ surface. 
To compare  the poloidal mode structure between the measurements and the synthetic data, we plot the \ac{SFL} angles of the channels using the $q=-5.35$ surface ($\Delta \Theta^{\star}_{q=-5.35}$) of the CLISTE equilibrium versus the phase determined from the sine fits in Fig.~\ref{fig:ECEphase}. The 'warm' resonance position is used to calculate $\Delta \Theta^{\star}_{q=-5.35}$ for channels with their cold resonance position in the optically thick (blue diamonds) and thin (green diamonds) region. The measured data agree very well with the synthetic data. The poloidal mode number of both is determined by fitting a linear function to the individual datasets (Fig.~\ref{fig:ECEphase}). From the slope of this linear function, we get $m_{ECE-I} ={\Delta \phi} / \Delta \Theta^{\star}_{q=-5.35} = 9.83 \pm 0.98$ using \ac{ECEI} and $m_{synth} = 10.72\pm 0.63 $ using synthetic data. 
Using only the \ac{ECEI} channels in the optically thick  does not strongly change  the result. There is a small difference between the prediction and the measurements, but it is within the uncertainties. Furthermore, one should also keep in mind that the $q$-profile contains also uncertainties, since the initial equilibrium calculation is constrained by measured density and temperature profiles.

The synthetic and the measured data indicate an almost resonant response at the $q=-5.35$ surface ($m_{pitch} \approx 10.7$). In fact, $m_{ECE-I}$ tends to be even lower than the pitch aligned mode number.
This is in contradiction to the non-resonant response with $|m|>|nq|$ expected from the magnetic perturbation. This seeming discrepancy originates from a difference in the poloidal mode structure between the magnetic perturbation and the flux surface displacement, which will be discussed in the following section. 

\subsection{Displacement versus magnetic perturbation}

\begin{figure*}
 \includegraphics[width=1.0\textwidth]{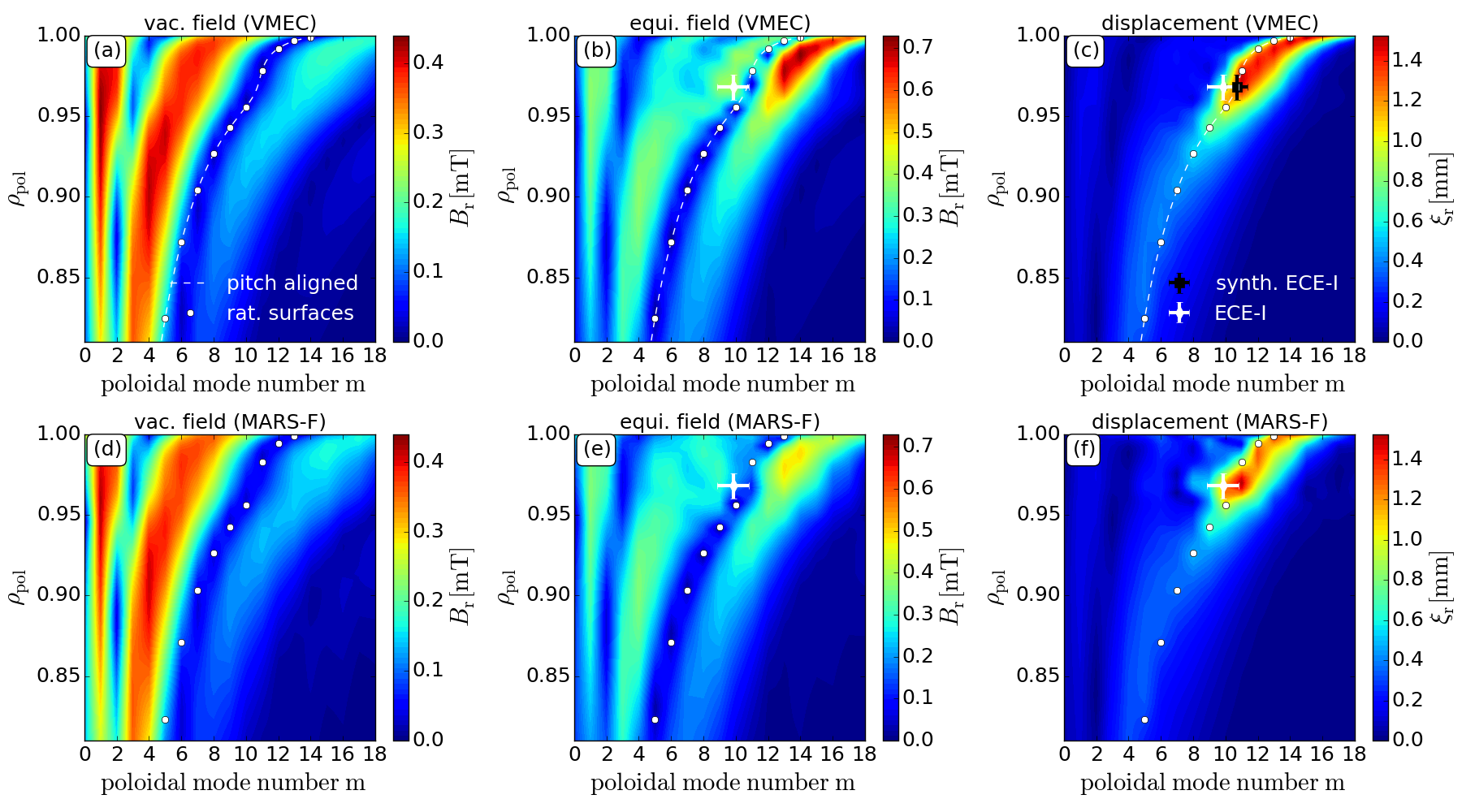} 
 \caption{The $n=-2$ poloidal mode spectrum of (a) the vacuum field perturbation, (b) the equilibrium field  perturbation, (c) the radial displacement from VMEC and (d) vacuum field perturbation (e) equilibrium field perturbation, (f) displacement from MARS-F. Note, there is no difference in the color scaling between VMEC and MARS-F. Poloidal mode number obtained from \ac{ECEI} is plotted (white diamond) as well as from the synthetic \ac{ECEI} (black square) in (c) to underline the consistency. The difference in the poloidal mode spectrum between the plasma response field and the displacement is apparent in both codes.}
\label{fig:poloidalAll}
\end{figure*}

To compare the flux surface displacement with the magnetic perturbation, we use a modified version of the MFBE code~\cite{Strumberger:1997} (described in \cite{Strumberger:2002}) to calculate the magnetic perturbation. In the following, we compare the mode spectra between the magnetic perturbation and the displacement predicted by VMEC and  MARS-F using ideal \ac{MHD}. For comparative purposes, Fig.~\ref{fig:poloidalAll}(a) and \ref{fig:poloidalAll}(d) show the vacuum field perturbation using the coordinate system from VMEC and MARS-F, respectively. The perturbation of the equilibrium field, which is the sum of the vacuum field and the plasma response field perturbation is plotted in Fig.~\ref{fig:poloidalAll}(b) and \ref{fig:poloidalAll}(e).

The equilibrium field perturbation predicted by VMEC as well as MARS-F is lowest around the resonant surfaces, which is expected from ideal \ac{MHD} (Fig.~\ref{fig:poloidalAll}(b) and (e)). Both calculations show a kink response situated at $|m|>|nq|$, which amplifies the field perturbation. This has been reported previously \cite{Ryan:2015, Lanctot:2011, Lanctot:2013, Haskey:2014} and has also been experimentally verified using probe measurements \cite{Lanctot:2011, King:2015}. On the contrary, the poloidal mode spectra of the flux surface displacement does not indicate such pronounced non-resonant components (Fig.~\ref{fig:poloidalAll}(c) and (f)). The structures are almost pitch aligned~\cite{Lazerson:2016}. The calculations from MARS-F and VMEC of the radial displacement  are in good agreement and they are also in-line with the \ac{ECEI} measurements (white diamond) within their uncertainties. 
The amplitudes of the displacement from both codes agree quantitatively, whereas the equilibrium field perturbations agree qualitatively.
Deviations from exactly zero resonant components at rational surfaces due to ideal \ac{MHD} can arise e.g. from numerical limitations and/or from the treatment of sheet currents on rational surfaces \cite{Reiman:2015}.  Detailed quantitative comparisons between MARS-F and VMEC are beyond the scope of this paper.

The analysis of the poloidal mode spectra of the flux surface displacement using \ac{ECEI} and its comparison to the 3D \ac{MHD} equilibrium codes is  relatively advanced. Its correctness relies on the accuracy of the individual steps like the modeling of the electron cyclotron radiation transport, the determination of the 'warm' resonance position,  the calculation of the  \ac{SFL} coordinate and thus, of the poloidal mode spectra, etc. Figure \ref{fig:poloidalAll}(c) shows the poloidal mode spectra of the displacement from VMEC. To demonstrate the consistency of the entire analysis chain, we also plot  the dominant poloidal mode number determined using the synthetic \ac{ECEI} diagnostic generated from the VMEC equilibrium (black rectangular) in Fig.~\ref{fig:poloidalAll}(c).  The point from the synthetic diagnostic overlies almost exactly the maxima of the poloidal mode spectra, which underlines the consistency and correctness of this analysis.

\subsection{Discussion of the poloidal mode structure}

While the calculated poloidal mode structure of the \ac{MP} is  non-resonant, the  measured displacement using \ac{ECEI} shows an almost resonant response. 
 This is a seeming contradiction, since the equilibrium field is the important parameter, which determines the displacement. Hence, one would expect that they have the same poloidal mode structure. But In the following, we will briefly show that this difference can already be explained by simple  ideal \ac{MHD} calculations.

In linear \ac{MHD}, the linearized magnetic perturbation $\vec{B}=\vec{B_0}+\vec{B_1}$  is related to the  surface  displacement $\vec{\xi}$ via~\cite{Freidberg:2014}:
\begin{equation} 
\label{equ:B1xi} 
\vec{B_1} = \vec{\nabla} \times (\vec{\xi} \times \vec{B_0}).
\end{equation} 
Assuming a cylindrical plasma ($r, \Theta, z$), the radial displacement $\xi_r$ and the radial magnetic field perturbation $B_r$  normal to the  axisymmetric flux surface relate:
\begin{equation} 
\label{equ:B1xi1a} 
B_r = \frac{B_{\Theta}}{r} \frac{\partial \xi_r}{\partial \Theta} + B_{\phi}   \frac{\partial \xi_r}{\partial z},
\end{equation} 
where $B_{\Theta}$ and $B_{\phi}$ are the poloidal and toroidal magnetic component, respectively. Using a periodic distortion $\xi \rm{_r}=\xi_a \ e^{i (m \Theta - \frac{n}{R} z )}$ and the resonant condition $q=\frac{m}{n}=\frac{r}{R}\frac{B_{\phi}}{B_{\Theta}}$, one gets the following relation at rational surfaces:


\begin{equation} 
\label{equ:B1xi2} 
B_r = i \frac{B_{\Theta}}{r} \left ( m - n q \right ) \xi_r \rightarrow \xi_r \propto \frac{B_{r}}{m - n q}
\end{equation} 
Consequently, $\xi_r$ maximizes at resonant surfaces in cylindrical plasmas. In the case of an elongated toroidal plasma, the elongated shape causes additional poloidal coupling between the different harmonics $\Delta m=0,1,2$. Then, the relation between $B_r$ and $\xi_r$ (Equ.~\ref{equ:B1xi2}) is not a relation between individual harmonics anymore. Nevertheless, the $ \frac{1}{m - n q}$ dependence in Equ.~\ref{equ:B1xi2}  is still the underlying reason for $\xi_r$  to be maximized at resonant surfaces, which is underlined by VMEC and MARS-F calculations (Fig.~\ref{fig:poloidalAll}(c,f)). This is also in-line with a new class of 3D ideal-MHD equilibria with nested surfaces and with current sheets at resonant surfaces producing a jump in the $q$-profile \cite{Loizu:2016}.

In summary, linear perturbative ideal \ac{MHD} calculations give a reasonable explanation for the differences in the mode structure between the surface displacement and magnetic perturbation. Initially, one idea of measuring the poloidal mode number was to distinguish resonant resistive \ac{MHD} response from non-resonant ideal \ac{MHD} response. Since this non-resonant ideal \ac{MHD} response appears practically resonant in the displacement, this method is not suitable to disentangle resistive from ideal \ac{MHD} response.

\section{Conclusions and summary}
\label{sec:conclusions}

The combination of a rigid rotating \ac{MP}-field and toroidally localized diagnostics provide a useful tool to measure the plasma surface distortion. This analysis relies on stable plasma conditions during the rigid rotation. \ac{ECE} diagnostics deliver informations about the plasma response via the flux surface displacement within the confined region, whereas \ac{ECE} measurements around the separatrix are difficult to interpret due to the transition from an optically thick to an optically thin plasma. 
Additional oblique angles of the \ac{LOS} complicate the interpretation of the \ac{ECE} data. It is therefore necessary to combine ray tracing with forward modeling of the radiation transport. The calculation of the 'warm' resonance positions (calculated maximum of the observed intensity distribution) is useful to estimate the real measurement position. The ideal \ac{MHD} equilibrium code VMEC was used to model the 3D plasma surface displacement and synthetic diagnostics were developed to compare the measurements with VMEC. 

A quantitative comparison of the displacement amplitude appears to be challenging. One can either use the single channel information  and/or the entire profile information for the comparison.  The first one is easier to realize, but implies the difficulties of  the large sensitivity on the measurement position and of the incomparability between the various profile diagnostics. Hence, we conclude that the use of the entire profile appears to be more useful. The comparisons of the displacement between synthetic and measured data are in reasonable agreement. The modeling underestimates only slightly the amplitude of the distortion on the \ac{LFS} midplane. Since MARS-F and VMEC exhibit very similar displacements, one plausible explanations for this minor underestimation could also be the uncertainties in the input parameters, like the pressure profile or the shape of the input equilibrium~\cite{Li:2016}. One should also keep in mind the role of the plasma position control during the rigid rotation. In the case of ASDEX Upgrade, the effect of the control system seems to be relatively small in the midplane due to the implementation of toroidal flux loops in the used reconstruction. In conclusion, not only the measurements of the amplitude require a careful treatment, but also the input parameters for the modeling in the presence of non-axisymmetric \acp{MP}.

The analysis of \ac{ECEI} shed some light onto the  plasma response  in terms of the flux surface displacement in the pedestal region and its poloidal mode structure. Differences in the poloidal mode structure  between the magnetic perturbations and the flux surface displacement are predicted by MARS-F and VMEC. The magnetic perturbation of the equilibrium field (vacuum field plus plasma response field) is non-resonant ($|m|>|nq|$), whereas the displacement is almost resonant ($|m|\approx|nq|$) as measured by \ac{ECEI} and as expected from ideal MHD in the vicinity of rational surfaces. Hence, it is not possible to use the poloidal mode number from the displacement to disentangle ideal from resistive \ac{MHD} response (resistive is always resonant).

The impact of the pressure profile and the differential phase angle $\Delta \phi \rm{_{ul}}$ on the displacement amplitude, although experimentally challenging, will be subject of future investigations.

\section{Acknowledgement}

The authors would like to thank V.~Igochine and E.~Wolfrum for fruitful discussions.
F.~M.~Laggner is a fellow of the Friedrich Schiedel Foundation for Energy Technology.
This work has been carried out within the framework of the EUROfusion Consortium and has received funding from the Euratom research and training programme 2014-2018 under grant agreement No 633053. The views and opinions expressed herein do not necessarily reflect those of the European Commission.

\bibliographystyle{unsrt}
\bibliography{plasmaresponse}

\end{document}